\documentclass[12pt]{iopart}
\usepackage{t1enc}

\def \Msub {{M}\hspace*{-.12in}\raisebox{.11in}{\em i}\hspace*{.05in}_{kj}}

\def \Msub0 {{M}\hspace*{-.12in}\raisebox{.11in}{\em i}\hspace*{.05in}_{j0}}
\def \Msub1 {{M}\hspace*{-.12in}\raisebox{.11in}{\em i}\hspace*{.05in}_{j1}}

\def \Mi {\stackrel{i}{M}}
\def \Mj {\stackrel{j}{M}}

\def \BE {\begin{equation}}
\def \EE {\end{equation}}
\def \BEAH {\begin{eqnarray*}}
\def \EEAH {\end{eqnarray*}}
\def \BEA {\begin{eqnarray}}
\def \EEA {\end{eqnarray}}
\def \BDM {\begin{displaymath}}
\def \EDM {\end{displaymath}}

\def \T {\Delta}

\def\bl {\bf{\ell}}
\def\vm {\bf{m}}
\def\vn {\bf{n}}
\def\bo {\cal{B}}
\def \bm {{\bf m}}
\def\vk {\bf{k}}

\def \bl {\mbox{\boldmath{$\ell$}}}

\def \btm {\mbox{\boldmath{$m$}}}
\def \cbm {\mbox{\boldmath{$\bar m$}}}
\newcommand{\lp}{\left(}
\newcommand{\rp}{\right)}

\usepackage{latexsym}
\newcommand{\qed}{%
  \ifmmode \quad\Box
  \else
    \leavevmode\unskip\penalty9999 \hbox{}\nobreak\hfill
    \quad\hbox{$\Box$}%
  \fi}

\newcommand{\fvec}[4]{{#1}_{\{a}{#2}_b{#3}_c{#4}_{d\}}}

\newcommand{\lnlm}[1]{\fvec{\ell}{n}{\ell}{m^{#1}}}
\newcommand{\lmlm}[2]{\fvec{\ell}{m^{#1}}{\,\ell}{m^{#2}}}

\newcommand{\mmlm}[3]{\fvec{m^{#1}}{m^{#2}}{\,\ell}{m^{#3}}}

\newcommand{\msub}[2]{m^{#1}{}_{#2}}
\newcommand{\msup}[2]{m_{#1}{}^{#2}}

\def \mS {\mbox{\boldmath{$S$}}}
\def \mA {\mbox{\boldmath{$A$}}}

\newcommand{\hfrac}[1]{#1\kern-1pt/\kern -.5pt2}

\newcommand{\authnote}[1]{}

\def \mS {\mbox{\boldmath{$S$}}}
\def \mA {\mbox{\boldmath{$A$}}}
\def \mL {\mbox{\boldmath{$L$}}}

\newcommand{\M}[3] {{\stackrel{#1}{M}}_{{#2}{#3}}}

\newcommand{\supth}{^{\mathrm{th}}}

\newcommand{\tT}{\mathbf{T}}

\newcommand{\AS}{\mathcal{A}}
\newcommand{\bV}{\mathbf{V}}

\newcommand{\eqref}[1]{(\ref{#1})}

\newtheorem{proposition}{Proposition}
\newtheorem{theorem}[proposition]{Theorem}

\begin{document}


\title{Classification of the Weyl Tensor in Higher Dimensions and Applications.}

\author{A. Coley\ddag}

\address{\ddag\ Department of Mathematics and Statistics, Dalhousie University,
Halifax, Nova Scotia, Canada}

\begin{abstract}

We review the theory of
alignment in Lorentz\-ian geometry and apply it to 
the algebraic classification of the Weyl tensor in
higher dimensions.
This classification reduces to the 
the well-known Petrov classification of
the Weyl tensor in four dimensions.
We discuss the  algebraic
classification of a number of known higher dimensional spacetimes.
There are many applications of the Weyl classification scheme,
especially when used in conjunction with the higher
dimensional frame formalism
that has been developed
in order to generalize the four dimensional
Newman--Penrose formalism. 
For example, we discuss
higher dimensional  generalizations of the Goldberg-Sachs
theorem and the peeling theorem. 
We also discuss the higher dimensional 
Lorentzian spacetimes with vanishing  scalar curvature
invariants and constant scalar
curvature invariants, which are of interest
since they are solutions of supergravity theory. 

\end{abstract}

\newpage
\section{Introduction}

We review the dimension-independent theory of
alignment, using the notions of an aligned null
direction and alignment order, in Lorentz\-ian geometry \cite{Algclass}. We then apply it to the tensor
classification problem for the Weyl tensor in higher dimensions \cite{class}.
In particular, it is
possible to categorize algebraically special tensors in terms of their
alignment type, with increasing specialization indicated by a higher
order of alignment.
In essence, one tries to normalize
the form of the tensor by choosing null vectors ${\bl}$ and ${\vn}$ so as to induce
the vanishing of the largest possible number of leading and trailing 
(with respect to boost weight)
null-frame scalars. The tensor can then be categorized by the extent
to which such a normalization is possible.
The approach reviewed is inspired by the work of Penrose and Rindler
used in general relativity \cite{penrind}, which is
reinterpreted and generalized in
terms of frame fixing.  The components of a Lor\-entz\-ian tensor can
be naturally ordered according to boost weight.  The approach is equivalent to
the well-known Petrov classification of
Weyl tensors in 4 dimensions \cite{petrov,Algclass}.

In practice, a complete tensor classification in terms of alignment type is possible only for
simple symmetry types and for small dimensions $N$ \cite{class}.  However, partial
classification into broader categories is still desirable.
We note that alignment type suffices for the classification of 4-dimensional
Weyl tensors, but
the situation for higher-dimensional Weyl tensors is more complicated.
For example, a 4-dimensional
Weyl tensor always possesses at least one aligned direction;
for higher dimensions, in general,
a Weyl tensor does not possess any aligned directions.
The end result of this approach is a
coarse classification, in the sense that it does not always lead to
canonical forms for higher-dimensional Weyl tensors
(unlike the 4-dimensional classification, in which each
alignment type admits a canonical form), and hence
is simply a necessary first step in the
investigation of covariant tensor properties.

In the higher dimensional classification, the secondary alignment type is
also of significance.  For algebraically special types, even though an aligned
null vector ${\bl}$ exists there does not,
generically, exist a second aligned null vector ${{\vn}}$, again in contrast to 4
dimensions.  Hence, in
higher dimensions we can distinguish algebraically special subclasses
that possess an aligned ${{\vn}}$.  Therefore, by definition, classification according to
alignment type corresponds to a normal form where the components of
leading and trailing boost weight vanish.  It would be desirable to
refine the classification to obtain   true canonical forms.
For example, this is possible in the case of the most algebraically special
Weyl tensors (e.g., type {\bf N} can always be
put into a canonical  form).
For more general Weyl tensors, the generic situation is that a number of
relevent Weyl tensor components (according to boost weight) are non-zero
which can be further simplified by performing spins and boosts (so that
hopefully the Weyl tensor can consequently be put
into a canonical form). Finally, the Ricci tensor can be further classified
according to its eigenvalue structure (i.e., Segr\'e type). It may also be of interest
to study the algebraic properties of the corresponding Weyl bivectors.

The classification outlined is a
general scheme which is applicable to the classification of arbitrary tensor
types in arbitrarily high dimensions.
Indeed, much of the analysis for higher
dimensional Weyl tensors can be applied directly to the classification of
higher dimensional Riemann curvature tensors.  In particular the
higher-dimensional alignment types
give well defined categories for the Riemann tensor (although there are additional constraints coming
from the extra non-vanishing components).
 The most important algebraic classification results in
4-dimensional general relativity is
the classification of the Weyl tensor according to Petrov type
and the classification of the Ricci tensor according to Segr\'e type.
We can also
use alignment to classify
the second-order symmetric Ricci tensor (which we refer to as Ricci type).
We can extend the classification of the Ricci tensor by describing a number of
additional algebraic types for the various alignment configurations and hence
obtain a complete classification; for example, by considering the geometric
properties of the corresponding alignment variety or by studying the Segr\'e type
\cite{Algclass}.

The classification of algebraic tensor types in Lorentzian geometry has a
purely mathematical importance. The indefinite signature makes
Lorentzian geometry profoundly different from Riemannian geometry.
The representation of the Weyl tensor is also of particular interest
from the physical standpoint.
First, as in four dimensions, tensor
classification is important for physical applications and, in
particular, for the study of exact solutions of the Einstein field equations
\cite{kramer}.  In addition, beyond the classical theory, there has been great
interest in higher dimensional Lorentz manifolds as models for
generalized field theories that incorporate gravity \cite{GSW,TOrtin}.

The main focus of the classification scheme is purely algebraic
(i.e.,  we consider tensors at a single point
of a Lorentzian manifold rather than tensor fields).
There is, however, a rich interplay between algebraic type and
differential identities. Indeed, as in 4 dimensions, in higher dimensions it is possible to use the Bianchi and Ricci equations to
construct algebraically special solutions of Einstein's field
equations, at least for the simplest algebraically special
spacetimes. The vast majority of currently known higher-dimensional
exact solutions are simple generalizations of 4-dimensional solutions. This
approach may lead to new, genuinely higher dimensional exact
solutions. Type {\bf N} and {\bf D} solutions may be of particular physical
interest.

\subsection{Overview}

In the next section we shall review the algebraic classification of the Weyl tensor in
higher dimensional Lorentzian manifolds,  by
characterizing algebraically special Weyl tensors by means of the
existence of aligned null vectors of various orders of alignment.
We shall discuss the principal classification in terms of the principal
type of the Weyl tensor in a Lorentzian manifold, and the
secondary classification in terms of the
alignment type.
For irreducible
representations the classification problem essentially reduces to the study of
the corresponding variety of all top alignment
schemes; we briefly discuss the alignment variety.
We discuss further classification in terms of reducibility, the dimension of the alignment variety, and
the multiplicity of principle directions. 
We then describe how the higher dimensional classification reduces to the classical
Petrov classification in 4 dimensions.

We next discuss the different
methods of classification, including a more practical method in terms of
necessary conditions. We then review the 
classification of a number of known higher dimensional spacetimes, including many
vacuum solutions of the Einstein field equations which
can represent higher dimensional black holes.

In section 3 we review a higher dimensional frame formalism that has been developed
in order to generalize the
Newman--Penrose formalism for any
$N>4$. In particular, we describe the
frame components of the Bianchi equations and the Ricci
identities.
There are many applications of the Weyl classification scheme,
especially when used in conjunction with the higher
dimensional frame formalism. For example, we shall discuss
higher dimensional  generalizations of the Goldberg-Sachs
theorem, the type-{\bf  D} conjecture
that asserts that stationary higher dimensional
black holes are  of Weyl type {\bf  D}, and the
peeling theorem. We also briefly discuss the five dimensional case
and type {\bf D} spacetimes.

In section 4 we shall  discuss the higher dimensional 
Lorentzian spacetimes with vanishing  scalar curvature
invariants ($VSI$ spacetimes) and constant scalar
curvature invariants ($CSI$ spacetimes).
Higher dimensional $VSI$ spacetimes are necessarily of Weyl type
{\bf III}, {\bf N} or {\bf O}, and higher dimensional $CSI$ spacetimes 
that are not locally homogeneous are
at most of type {\bf II}. We begin with a discussion of
higher dimensional Kundt spacetimes and  spacetimes that admit a
covariantly constant null vector (which are Kundt and 
generically of Ricci and Weyl type {\bf II}).

All of the higher dimensional $VSI$ spacetimes are known
explicitly. Various classes of $CSI$ spacetimes are discussed, and it has
been conjectured that if a spacetime
is $CSI$ then that spacetime is either locally homogeneous or it
belongs to the higher dimensional Kundt class and  can be constructed from locally
homogeneous spaces and $VSI$ spacetimes. The 
various $CSI$ conjectures are discussed. 
The $VSI$ and $CSI$ spacetimes are of particular interest
since they are solutions of supergravity or superstring theory, when supported by
appropriate bosonic fields. 
It is known that the higher-dimensional 
$VSI$ spacetimes with fluxes and dilaton are solutions of type IIB
supergravity, and it has been argued that they
are exact solutions in string theory.
$VSI$ solutions of IIB supergravity with NS-NS and
RR fluxes and dilaton have been explicitly constructed.
We discuss 
$CSI$ spacetimes that are solutions of supergravity.
The supersymmetry properties of the  $VSI$ and $CSI$
spacetimes are also discussed. 

In the final section we briefly discuss some of the 
outstanding problems and outline possible future work.

\newpage

\section{Classification of the Weyl Tensor in Higher Dimensions}

Let us review the algebraic classification of the Weyl tensor in
higher N-dimensional Lorentzian manifolds \cite{class,Algclass},  by
characterizing algebraically special Weyl tensors by means of the
existence of aligned null vectors of various orders of alignment.
Further classification is obtained by specifying the alignment
type and utilizing the notion of reducibility.
The present classification reduces to the classical
Petrov classification in 4 dimensions \cite{petrov}.

We shall consider a null frame  ${\bl}=\vm_0,\;{{\vn}}=\vm_1,\;\vm_2,...\vm_i$
(${\bl},\;\vn$ null with  $\ell^a \ell_a = n^a n_a = 0, \ell^a n_a = 1$,
$\vm^i$ real and spacelike with $m_i{}^a m_j{}_a = \delta_{ij}$; all other products vanish)
in an  $N$-dimensional Lorentzian spacetime with signature $(-+...+)$,
so that $g_{ab} = 2l_{(a}n_{b)} +  \delta_{jk} m^j_a m^k_b$.
Indices $a,b,c$ range from $0$ to $N-1$, and
 space-like indices $i,j,k$ also indicate a
null-frame, but vary from $2$ to $N-1$ only.
We note that all notation and conventions (e.g., for the definitions
of the Riemann and Ricci tensors) follow that of
\cite{Algclass,class,Bianchi}.

The frame is covariant relative to the group of {\em real} orthochronous linear
Lorentz transformations, generated by null rotations, boosts and spins.
A {\em null rotation} about $\vn$ is a
Lorentz transformation of the form
\begin{equation}
  \label{eq:nullrot}
    \hat{\vn}=  \vn,\quad
    \hat{\vm}_i=  \vm_i + z_i \vn,\quad
    \hat{{\bl}}= \textstyle {\bl}
    -z_i \vm^i-\frac{1}{2} \delta^{ij} z_i z_j\, \vn.
\end{equation}
A null rotation about ${\bl}$ has an analogous form.
A spin is a transformation of the form
\begin{equation}
  \label{eq:spin}
    \hat{\vn}= \vn, \quad
    \hat{\vm}_i=  X_i^j\,\vm_j,\quad
    \hat{{\bl}}=  {\bl},
\end{equation}
where $X_i^j$ is an orthogonal matrix. Finally, and most
importantly, a {\em boost} is a transformation of the form
\begin{equation}
  \label{eq:boost}
    \hat{\vn}= \lambda^{-1}\vn,\quad
    \hat{\vm}_i=  \vm_i,\quad
    \hat{{\bl}}=  \lambda\, {\bl}, \quad \lambda \neq 0.
\end{equation}

Let $T_{a_1... a_p}$ be a rank $p$ tensor.  For a fixed list of
indices $A_1,...,A_p$, we call the corresponding $T_{A_1... A_p}$
a null-frame scalar.  These scalars transform under a boost
(\ref{eq:boost}) according to
\begin{equation}
  \label{eq:boostxform}
  \hat{T}_{A_1... A_p}= \lambda^b\,
  T_{A_1... A_p},\quad b=b_{A_1}+...+b_{A_p},
\end{equation}
where $b_0=1,\quad b_i=0,\quad b_1=-1$. We call $b$ the
boost-weight of the scalar. We define the {\em boost order} of the
tensor $T$ to be the boost weight of its leading term (i.e., the
maximum value for  $b_{A_1... A_p}$ for all non-vanishing
$T_{A_1... A_p}$). The result of applying null rotations, boosts and spins on
the components of the Riemann, Ricci and Weyl tensors and the Ricci
rotation coefficients are given in \cite{Algclass,Bianchi,ricci}.

We introduce the notation \BE
T_{\{pqrs\}}\equiv\frac{1}{2}(T_{[pq][rs]}+
 T_{[rs][pq]}). \EE
We can decompose the Weyl tensor and sort the components of the
Weyl tensor by boost weight \cite{Algclass}:

\begin{eqnarray}
 C_{abcd} = {\overbrace{4C_{0i0j} n_{\{a}{m^i}_b n_c {m^j}_{d\}}
}}^2 + {\overbrace{8C_{010i} n_{\{a} \ell_b n_c {m^i}_{d\}}
 + 4C_{0ijk} n_{\{a} {m^i}_b {m^j}_c {m^k}_d\}}}^1 + && \nonumber \\
 \left\{\begin{array}{l}
 4C_{0101} n_{\{a} \ell_b n_c \ell_{d\}} + 4C_{01ij}
 n_{\{a} \ell_b {m^i}_c {m^j}_{d\}} +\label{eqnweyl} \\
 8C_{0i1j} n_{\{a} {m^i}_b \ell_c {m^j}_{d \}} +
 C_{ijkl} m^i_{\{a} {m^j}_b  {m^k}_c {m^l}_{d\}}\end{array}
 \right\}^0 + && \\
 {\overbrace{8C_{101i} \ell_{\{a} {n}_b \ell
 _c {m^i}_{d\}} + 4C_{1_{ijk}} \ell_{\{a} {m^i}_b {m^j}_c {m^k}_{d \}}}}^{-1}
 + {\overbrace{4C_{1i1j} \ell_{\{ a} {m^i}_b \ell_c {m^j}_{d\}}}}^{-2}.&& \nonumber
 \end{eqnarray}

\begin{table}[htbp]
  \begin{center}
    \begin{tabular}{|c|c|c|c|c|}
      \hline
      $2$ & $1$ & $0$ & $-1$ & $-2$\\
      \hline
      $C_{0i0j}$ &  $C_{010i}, C_{0ijk}$ & $C_{0101}, C_{01ij},
      C_{0i1j}, C_{ijkl}$ & $C_{011i}, C_{1ijk}$ & $C_{1i1j}$\\
      \hline
    \end{tabular}
    \medskip
    \caption{Boost weights of  the Weyl scalars.}
    \label{tab:bw2}
  \end{center}
\end{table}

The Weyl  tensor is generically of boost order
$2$. If all $C_{0i0j}$ vanish, but some $C_{010i}$, or $C_{0ijk}$
do not, then the boost order is $1$, etc. The Weyl scalars also
satisfy a number of additional relations, which follow from
curvature tensor symmetries and from the trace-free condition:
\begin{eqnarray}
C_{0i0}{}^i = 0, C_{010j} = C_{0ij}{}^i,  C_{0(ijk)} = 0, C_{0101}
= C_{0i1}{}^i,C_{i(jkl)} = 0,\nonumber\\
C_{0i1j}=-\frac{1}{2} C_{ikj}{}^k+\frac{1}{2}C_{01ij},  C_{011j} = -C_{1ij}{}^i,  C_{1(ijk)} = 0,  C_{1i1}{}^i =
0\label{cons}.
\end{eqnarray}
A real null rotation about ${\bl}$ fixes the leading terms of a tensor,
while boosts and spins subject the leading terms to an invertible
transformation.  It follows that the boost order (along ${\bl}$) of a tensor is a
function of the null direction ${\bl}$ (only). We shall therefore
denote boost order by $\bo({\bl})$ \cite{Algclass}. We {\em define}
a null vector ${\bl}$ to be {\em aligned} with the Weyl tensor
whenever $\bo({\bl})\leq$ $1$ (and we shall refer to ${\bl}$ as a Weyl
aligned null direction (WAND)). We call the integer
$1-\bo({\bl})\in$ $\{0,1,2,3\}$ the order of alignment. The alignment
equations are $\frac{1}{2}N(N-3)$ degree-4 polynomial
equations in $(N-2)$ variables, which are in general overdetermined
and hence have no solutions for $N>4$.

Any tensor, including the Riemann tensor, can be algebraically classified
according to boost weight in arbitrary dimensions in a
similar way \cite{Algclass}. However, the value of the maximum boost order will
depend on the rank and symmetry properties of the tensor.
In particular, we can also
use alignment to classify
the second-order symmetric Ricci tensor in higher
dimensions.
We can extend the classification of the Ricci tensor by describing a number of
additional algebraic types for the various alignment configurations and hence
obtain a complete classification (for example, by considering the geometric
properties of the corresponding alignment variety or by studying the Segr\'e type
\cite{Algclass}).

\subsection{Principal Classification:}

Following \cite{Algclass}, we say that the {\bf principal
type} of the Weyl tensor in a Lorentzian manifold is {\bf I, II,
III, N} according to whether there exists an aligned ${\bl}$ of
alignment order $0,1,2,3$ (i.e. $\bo({\bl})=$ $1,0,-1,-2$),
respectively. If no aligned ${\bl}$ exists we will say that the
manifold is of (general) type {\bf G}. If the Weyl tensor
vanishes, we will say that the manifold is of type {\bf O}. The
algebraically special types are summarized as follows:
\begin{eqnarray}
&&  Type ~~{\bf I}: ~~ C_{0i0j}=0 \nonumber\\
&&  Type~~ {\bf II}: ~~C_{0i0j}=C_{0ijk}=0 \nonumber \\
&&  Type ~~{\bf III}:  ~~C_{0i0j}=C_{0ijk}=C_{ijkl} =C_{01ij}=0 \nonumber\\
&&  Type ~~{\bf N}:  ~~C_{0i0j}=C_{0ijk}=C_{ijkl} = C_{01ij}=C_{1ijk}=0
\end{eqnarray}

Note that the number of independent frame components of various
boost weights is \BDM \overbrace{2 \lp \frac{N(N-3)}{2}
\rp}^{2,-2} + \overbrace{2 \lp \frac{(N-1)(N-2)(N-3)}{3} \rp}^{1,-1}
+\overbrace{\frac{(N-2)^2(N^2-4N-5)}{12} + \frac{(N-2)(N-3)}{2}}^0 , \EDM
where the number of independent
components of the Weyl tensor is\BDM
  \label{eq:wdim}
  \frac{(N+2)(N+1)N(N-3)}{12}  .
\EDM
A 4-dimensional
Weyl tensor always possesses at least one aligned direction.
For higher dimensions, in general,
a Weyl tensor does not possess any aligned directions.
Indeed, in \cite{Algclass} it was shown that
if $N\geq 5$, then the set of Weyl tensors with alignment type
{\bf  G} is a dense, open subset of the set (vector space) of all
$N$-dimensional Weyl tensors.

\subsection{Secondary Classification:}

Further categorization can be obtained by
specifying {\em alignment type} \cite{Algclass}, whereby we try to
normalize the form of the Weyl tensor by choosing both ${\bl}$ and
${\vn}$ in order to set the maximum number of leading and trailing
null frame scalars to zero. Let ${\bl}$ be a WAND whose order of
alignment is as large as possible. We then define the principal (or primary)
alignment type of the tensor to be $b_{max} - b({\bl})$. Supposing
such a WAND ${\bl}$ exists, we then let $\vn$ be a null vector of
maximal alignment subject to $\ell_{a} n^{a}=1$. We define the
secondary alignment type of the tensor to be $b_{max}-b(\vn)$. The
alignment type of the Weyl tensor is then the pair consisting of
the principal and secondary alignment type
 \cite{Algclass}. In general, for types ${\bf I}, {\bf II},
{\bf III}$ there does not exist a secondary aligned $\vn$ (in
contrast to the situation in 4 dimensions), whence the
alignment type consists solely of the principal alignment type.
Alignment types (1,1), (2,1) and (3,1) therefore form
algebraically special subclasses of types ${\bf I}, {\bf II}, {\bf
III}$, respectively (denoted ${\bf I}_i, {\bf II}_i, {\bf
III}_i$). There is one final subclass possible, namely type (2,2)
which is a further specialization of type (2,1); we shall denote
this as type ${\bf II}_{ii}$ or simply as type ${\bf D}$.
Therefore, a type ${\bf D}$ Weyl tensor in canonical form has no
frame components of boost weights $2,1,-1,-2$ 
(i.e., all frame components are of boost
weight zero).\\

\newpage

\subsection{The alignment variety}

In the classification scheme we essentially study the
locus of aligned null directions, where the
singular points correspond to null directions of higher
alignment (for which multiple leading orders vanish).
For irreducible
representations the classification problem is reduced to the study of
the corresponding variety of all top alignment
schemes quotiented by the group of M\"obius transformations.
Let us discuss this
in more mathematical detail.

It can be shown that
the set of null directions aligned with a fixed $N$-dimensional tensor
is a variety (called the {\em top alignment variety}); i.e.,  it
can be described by a certain set of polynomial equations in $N-2$
variables \cite{Algclass}.  These
alignment equations are given with respect to a particular null-frame.
A change of null-frame transforms the equations in a covariant
fashion.  The necessary mathematical framework needed to describe such
covariant compatibility is a scheme.

The null directions of higher alignment order are subvarieties of the
top alignment variety.  These higher-order directions have an important
geometric characterization related to singularities.  Indeed, it can be proven
that for irreducible representations of the Lorentz group, the
equations for higher order alignment are equivalent to the equations
for the singular points of the top alignment variety \cite{Algclass}.  Thus, for the Weyl
tensor, the
algebraically special tensors can be characterized by the singularities of the
corresponding alignment variety.

Indeed, it can be shown that the set of aligned directions is a
M\"obius variety,
by exhibiting the compatible equations for this set.
By definition, a null vector
$$\vk = {\bl} -\frac{1}{2} z^iz_i \vn + z^i \vm_i,$$
is aligned with a rank $p$ tensor $\tT$, with alignment
order $q$, if the corresponding
$q\supth$ order \emph{alignment equations} (polynomial in $z_i$) are satisfied.
The ideal generated by the above polynomials is called
the \emph{alignment ideal} of order $q$.
The alignment scheme of order $q$, $\AS^q=\AS^q(\tT)$, is
  the scheme generated by the alignment ideals $A^q(\tT)$.  The top
  alignment scheme, denoted $\AS\equiv\AS^0(\tT)$ ($q=0$),
  has a distinguished role.
The corresponding varieties $\bV(\AS^q)$ consist of aligned null
directions of alignment order $q$ or more.  It is possible that
the alignment equations are over-determined and do not admit a
solution, in which case the variety is the empty set.
In those cases
where $\bV(\AS^q)$ consists of only a finite number of null
directions, we call these directions \emph{principal} and speak
of the \emph{principal null directions} (PNDs) of the tensor.

A point of the variety $z_i\in\bV(I)$ will be called
\emph{singular} if the first-order partial derivatives of the polynomials in
the affine ideal $I$ also
vanish at $z_i$.  It will be called singular of order $q$ if all partials of
order $q$ and lower vanish at that point. Geometrically, a singular point
represents a self-intersection such as a a node or a cusp, or a point of higher
multiplicity.  It was shown \cite{Algclass} that if $\vk$ spans an aligned direction of order
$q$, then $[\vk]$ is a $q\supth$ order singular point of the top variety $\bV(\AS)$.
Furthermore, if $\tT$ belongs to an irreducible representation of the Lorentz group
and $\vk$ is a $q\supth$-order singular element of the top alignment variety, then
$\vk$ spans a $q\supth$ order aligned null direction (and consequently the $q\supth$
order alignment scheme $\AS^q$ describes the $q\supth$-order singular points of the
top scheme $\AS$).

\newpage

\section{Further classification}

\subsection{Decomposability and Reducibility}

The Weyl tensor $C_{abcd}$ is reducible if and only if it is the sum of two Weyl
tensors, one of which is a Weyl tensor of an (irreducible)
Lorentzian space of dimension $M$, and the other is a Weyl tensor
of a Riemannian space of dimension $N-M$.  It turns out that
subclassification of the Weyl tensor is most easily accomplished
by decomposing the Weyl tensor and classifying its irreducible
parts.

Let us consider a spacetime, $M$, which is a paracompact, Hausdorff,
simply connected smooth Lorentzian manifold of arbitrary
dimension $N$.  We can discuss the local
decomposability and reducibility of manifolds in terms of
holonomy groups and their associated holonomy algebras.
Let us work in local coordinates, where lower case Latin
indices range from $0, \ldots, N-1$, and Greek and uppercase
Latin indices have the ranges $\alpha, \beta, \gamma = 0, \ldots, M-1$ and $A,B,C
= M, \ldots, N-1$, respectively. We consider the Lorentzian manifold $M$ to
be an N-dimensional product manifold:

\begin{equation}
M_N = M_M \otimes \tilde{M}_{\tilde{M}} ; N = M+ \tilde{M},
\end{equation}
where $M_M$ is an M-dimensional Lorentzian manifold and
$\tilde{M}_{\tilde{M}}$ an ${\tilde{M}}$-dimensional Riemannian
manifold, with metric
\begin{equation}
\label{metric}
 g_N = g_M \oplus \tilde{g}_{\tilde{M}},
\end{equation}
or in coordinates
\begin{equation}
\label{metriccomps} ds^2 = g_{\alpha\beta}(x^\gamma) dx^\alpha
dx^\beta + \tilde{g}_{AB} (y^C) dy^A dy^B.
\end{equation}

A product space $M_N = M_M \otimes \tilde{M}_{\tilde{M}}$ is said
to be {\em decomposable} and the metric tensor $g$ can be written in
terms of  (\ref{metric}) and (\ref{metriccomps}). An object in a
decomposable $M_N$ is called breakable if its components with
respect to the coordinates defined in (\ref{metriccomps}) are
always zero when they have indices from both ranges.  A breakable
object is said to be decomposable if and only if the components
belonging to $M_M$ (resp., $\tilde{M}_{\tilde{M}}$) depend on
$x^\gamma$ (resp., $y^C$) only.  The Riemann tensor in a
decomposable manifold is decomposable  (it follows immediately
that the Riemann tensor is breakable \cite{Beem}, and if the
Riemann tensor is breakable it follows from the Bianchi identities
that it is decomposable). Symbolically this can be written as:
\begin{equation}
R_N = R_M \oplus \tilde{R}_{\tilde{M}}; ~~ R^{a}_{~bcd} =
R^{\alpha}_{~\beta\gamma\delta} + {\tilde R}^{A}_{~BCD}.
\end{equation}
Alternatively, we can write the Riemann tensor
in the suggestive block diagonal form: $R^{a}_{~bcd} = {\bf block
diag}\{ R^{\alpha}_{~\beta\gamma\delta}(x^\epsilon) ,{\tilde
R}^{A}_{~BCD}(y^E)\}$.
Since the manifold is a product space, if $R^{\alpha}_{~\beta \gamma
\delta}$ is decomposable, then so is $R_{\alpha \beta \gamma \delta}$.
This can be trivially generalized to:
\begin{equation}
M_N = M_M \otimes \tilde{M}_{\tilde{M}_1} \otimes
\tilde{M}_{\tilde{M}_2} \otimes \ldots; N = M + \tilde{M}_1 +
\tilde{M}_1 + \ldots; ~~ g_N = g_M \oplus \tilde{g}_{\tilde{M}_1}
\oplus \tilde{g}_{\tilde{M}_2} \oplus \ldots
\end{equation}

The connection in a decomposable manifold is also decomposable,
and each of $M_M$ and $\tilde{M}_{\tilde{M}}$ are consequently
totally geodesic \cite{Beem,Schouten,O'Neill}. Conversely, if a
manifold $M$ is constructed from two totally geodesic submanifolds
$M_M$ and $\tilde{M}_{\tilde{M}}$, it is decomposable. In
addition, if a manifold is decomposable, then there exists a real
non-trivial covariantly constant geometrical field (not proportional to the
metric $g$).  Such a covariantly constant field is necessarily
invariant under the holonomy group, and hence decomposable
manifolds are associated with manifolds with a reducible holonomy
group.  The generators of the holonomy group span the decomposable
parts of the Riemann tensor (and its covariant derivatives).

$M$ is said to be {\em reducible} if the holonomy group is reducible as
a linear group acting on the tangent bundle (otherwise it is said to be
irreducible). Let $M$ be reducible and let $T^\prime$ denote the
distribution associated with the proper (non-trivial) subspace
invariant under the holonomy group.  Then $T^\prime$ is
differentiable and involutive and the maximal integral manifold
$M^\prime$ of $T^\prime$ through a point of $M$ is a totally
geodesic submanifold of $M$ \cite{KN}.  In addition, if
$T^{\prime\prime}$ is a complementary and orthogonal
distribution to $T^\prime$ at each point of $M$, and $M^\prime$
and $M^{\prime\prime}$ are their associated maximal integral manifolds, then
each point $p$ of $M$ has an open neighbourhood $V= V^\prime
\times V^{\prime \prime}$, where $V^\prime$ (resp., $V^{\prime
\prime}$) is an open neighbourhood of $p$ in $M^\prime$ (resp.,
$M^{\prime\prime}$), and the metric in $V$ is the direct
product of the metrics in $V^\prime$ and $V^{\prime\prime}$
\cite{KN}.  The converse is also true \cite{Beem}.    The de Rham
decomposition theorem then asserts that if $M$ is connected,
simply connected and complete then $M$ is isometric to the direct
product $M^\prime \otimes M^{\prime\prime}$ \cite{KN, Geroch}.

The case $N=4$ has been studied in detail. 
A 4-dimensional Lorentzian spacetime $M$ is reducible if the holonomy
group of $M$ is reducible; i.e., the holonomy algebra of $M$ is a
proper subalgebra of the Lorentz algebra and of (holonomy) type
$R_2 - R_{14}$ $(R_1$ is the trivial case, $R_{15}$ is the full
Lorentz algebra) \cite{Hall}. Therefore, the holonomy group is reducible in
all cases except for the general type $R_{15}$, and $M$
consequently has a product structure in all cases where the
reduction is non-degenerate \cite{Beem}. If the Riemann tensor is
decomposable, then the 4-dimensional spacetime is $1+3$ or $2+2$
decomposable.  Let $x^\alpha$ and $y^A$ be the associated local
coordinates. Suppose that $R = R(x^\alpha)
 \oplus \tilde{R}(y^A)$. Then there is a product manifold with metric
$g(x^\alpha) \oplus  \tilde{g}(y^A)$ where $R(x^\alpha)$ and
$\tilde{R}(y^A)$ are the Riemann tensors associated with
$g(x^\alpha)$ and $\tilde{g}(y^A)$, respectively. Suppose that
$\overline{g}$ is also a metric with Riemann tensor $R$. Then
$\overline{g}$ is explicitly related to $g$ in terms of
eigenvectors that live in $M$ or $\tilde{M}$ \cite{Hall}. The
metric $\overline{g}$ still has a product structure, but
$\overline{g}$ is not necessarily isometric to $g$ (i.e., although
all metrics compatible with $R$ are products, the metric is not
unique).

\subsubsection{The Weyl tensor:}

In the case in which the Weyl tensor is reducible, it is possible
to obtain more information by decomposing the Weyl tensor and algebraically
classifying its irreducible parts. In the case of vacuum, it
follows from above that if $C^{a}_{~bcd}$ is breakable, it is also
decomposable and the manifold has a product structure. In general,
it does not follow that if $C^{a}_{~bcd}$ is breakable then it is
also decomposable. For a product space manifold, if
$C^{a}_{~bcd}$ is decomposable then so is $C_{a bcd}$. However, in
general the manifold is not a product space and hence the
decomposibility of $C_{a bcd}$ is not equivalent to the
decomposibility of $C^{a}_{~bcd}$.

The Weyl tensor $C_{abcd}$ is {\em reducible} if there exists a
null frame and a constant $M<N$  such
that:
\begin{equation}
 C_N = C_M \oplus \tilde{C}_{\tilde{M}}; ~~
C^{a}_{~bcd} = C^{\alpha}_{~\beta\gamma\delta} + {\tilde
C}^{A}_{~BCD}.\label{weyl}
\end{equation}
If the Weyl tensor is (orthogonally) reducible, the only non-vanishing components
of ${^N}C_{abcd}$ are
$$C_{\scriptscriptstyle \alpha\beta\gamma\delta} \neq 0,\quad 0\leq \alpha,\beta,\gamma,\delta\leq M-1;\quad
{\tilde{C}}_{\scriptscriptstyle ABCD} \neq 0
  \quad M\leq A,B,C,D\leq N-1$$
(i.e., in a fixed frame, there are no Weyl scalars  ${^N}C_{abcd}$
of mixed type).
Therefore, $C_{abcd}$ is (algebraically)
reducible if and only if it is the sum of two Weyl tensors
(and the \emph{reduced alignment type} of $C_{\alpha \beta \gamma \delta}$
can be defined to be the
alignment type of the reduced tensor). A
reducible Weyl tensor is said to be (geometrically) decomposable
if and only if the components belonging to $M_M$ (resp.,
$\tilde{M}_{\tilde{M}}$) depend on only $x^\gamma$ (resp., $y^{C}$)
\cite{KN}: ${^M}C_{\alpha\beta\gamma\delta}$ is a Weyl tensor of
an ({\em irreducible}) Lorentzian spacetime of dimension $M$,
${{^{\tilde M}}{\tilde C}}_{\scriptscriptstyle ABCD}$ is a Weyl tensor of a Riemannian
space of dimension $\tilde{M} \equiv N - M$.
Note that non-trivial Weyl tensors do not exist if
$N<4$.  Thus, reducibility becomes an issue only in dimensions $N\geq
5$, and even then one of the complementary summands must necessarily
vanish for $N\leq 7$.  Therefore, true decomposability can only occur for $N\geq
8$.

We can also write a decomposable Weyl
tensor in the suggestive block diagonal form: ${\bf block diag}\{
C_{\alpha\beta\gamma\delta}(x^\epsilon) ,{{C}}_{\scriptscriptstyle
ABCD}(y^{E})\}$. This can be trivially generalized to the case
where $\tilde{M}_{\tilde{M}}$ is further reducible. Writing out
the Weyl tensor in terms of boost weights we obtain
(\ref{eqnweyl}), with indices $\alpha,\beta,\gamma,..$ running
from $0 - (M-1)$, and an additional term
\begin{equation}
(+) ~~~ \{{C}_{IJKL} m^{I}_{\{A}m^{J}\ \!_{B} m^{K}\ \!_{C} m^{L}\
\!_{D\}} \}^0.
\end{equation}
This term, corresponding to the Riemannian part of the Weyl tensor
${\tilde C}$, is either identically zero (of type ${\bf O}$) or
has terms of boost weight zero only and hence is of type ${\bf
D}$.

We note that the Weyl tensor is clearly reducible for a manifold that is
conformal to a product manifold. In particular, the reducibility
property for the Weyl tensor applies to a warped product manifold
with metric:
\begin{equation}
ds^2 = g_{\alpha\beta} (x^\gamma) dx^\alpha dx^\beta + F(x^\gamma)
\tilde{g}_{AB} (y^C) dy^A dy^B.
\end{equation}
This is conformal to the metric
\begin{equation}
ds^2 = F(x^\gamma) \{F^{-1}(x^\gamma) g_{\alpha\beta} (x^\gamma)
dx^\alpha dx^\beta + \tilde{g}_{AB} (y^C)dy^A dy^B\},
\end{equation}
and hence the two conformally related manifolds have the same Weyl
tensor; i.e., the warped manifold has the same Weyl tensor as the
conformally related product manifold.
A conformally decomposable manifold (doubly warped or
twisted manifold) is of the form:
\begin{equation} ds^2 = F(x^\gamma,y^C)
\{g_{\alpha\beta} (x^\gamma) dx^\alpha dx^\beta\} + {\tilde
F}(x^\gamma,y^C)\{\tilde{g}_{AB} (y^C)dy^A dy^B\}
\end{equation}
(which is different to a manifold that is conformal to a product
manifold).

We note that almost all higher dimensional manifolds of physical
interest are either product or warped product manifolds \cite{RT}.
Let us assume that the Weyl tensor is reducible as in
(\ref{weyl}), where ${^M}C$ and $\tilde C$  are the $M$-
dimensional irreducible Lorentzian and $(N-M)$- dimensional
Riemannian parts. Then associated with each part would be a
principal type. The principal type of the Lorentzian ${^M}C$ would
be ${\bf G},{\bf I},{\bf II},{\bf III},{\bf N}$, or ${\bf O}$. However,
the principal type of the Riemannian $\tilde C$ is either ${\bf
D}$  or ${\bf O}$. We could denote a secondary type of a
reducible Weyl tensor (\ref{weyl}) as $T_{M} \times {\tilde
T}_{\tilde M}$, or simply by $T_{\tilde T}$ if the dimensions $M,
{\tilde M}$ are clear. For example, ${\bf I}_{\bf D}$ would denote
a reducible Weyl tensor in which the irreducible Lorentzian part
of the Weyl tensor is of principal type ${\bf I}$, and the
irreducible Riemannian part of the Weyl tensor is non-zero.

\subsection{Full Classification:}

Alignment type by itself is insufficient for a complete
classification of the Weyl tensor. It is necessary to count
aligned directions, the dimension of the alignment variety, and
the multiplicity of principle directions. We note that unlike in
4 dimensions, it is possible to have an infinity of
aligned directions. If a WAND is discrete, for consistency with
4-dimensional nomenclature we shall refer to it as a principal
null direction (PND). We can introduce extra normalizations and
obtain further subclasses.
Complications
in attempting to find canonical forms arise due to gauge fixing
(i.e., certain terms can be chosen to be zero by an appropriate
choice of frame through boosts and spatial rotations).

The {\bf full type} of an irreducible Weyl tensor $C_{abcd}$ is
defined by its principal and secondary types, and includes all of
the information on subcases and multiplicities. We do not classify
these in full detail here (indeed, it may be necessary
to consider different specific dimensions on a case by case basis),
but rather describe some of the key
algebraically special subtypes below.

First, there are additional conditions for algebraic
specializations: (i) Type {\bf I} (a)  $C_{010i}=0$. (ii) 
Type {\bf II} (a) $C_{0101}=0$, (b) the traceless Ricci part of
$C_{ijkl} = 0$, (c) the Weyl part of $C_{ijkl} = 0$, (d) $C_{01ij}
= 0$. (iii) Type {\bf III} (a) $C_{011i}=0$. For example, there are
two subcases of type {\bf III}, namely type {\bf III} (general
type) and type {\bf III} (a) in which $C_{011i}=0$.
Second, there are further specializations due to multiplicities.
In types {\bf III} and {\bf N} all WANDs are necessarily PNDs. For
type {\bf III} tensors, the PND of order 2 is unique. There are no
PNDs of order 1, and at most 1 PND of order 0.  For $N=4$ there is
always exactly $1$ PND of order 0. For $N>4$ this PND need not
exist. For type {\bf N} tensors, the order 3 PND is the only PND
of any order.

We can write a canonical form for Weyl type {\bf N}.
From (\ref{eqnweyl})--(8), we have that for type {\bf N}:
\begin{equation}
C_{abcd} = 4C_{1i1j} \ell_{\{ a} {m^i}_b \ell_c {m^j}_{d\} };
~~C_{1i1}{}^i = 0.
\end{equation}
The general form of the Weyl tensor for type {\bf III} is given by
(\ref{eqnweyl}) subject to (\ref{cons}) and (8). In the subclass {\bf III}$_i$
we have that $C_{1i1j}=0$ and hence
\begin{equation}
C_{abcd} = 8C_{101i} \ell_{\{a} {n}_b \ell
 _c {m^i}_{d\}} + 4C_{1ijk} \ell_{\{a} {m^i}_b {m^j}_c {m^k}_{d
 \}},
\end{equation}
where $C_{011j} = -C_{1ij}{}^i, C_{1(ijk)} = 0$, and not all of
$C_{1ijk}$ are zero (else it reduces to Weyl type ${\bf O}$). In the
subclass {\bf III}(a) we have that $C_{011j} = 0$, so that
\begin{equation}
C_{abcd} = 4C_{1_{ijk}} \ell_{\{a} {m^i}_b {m^j}_c {m^k}_{d
 \}} + 4C_{1i1j} \ell_{\{ a} {m^i}_b \ell_c {m^j}_{d\} },
\end{equation}
where $C_{1ij}{}^i = C_{1(ijk)} = C_{1i1}{}^i = 0$. There is a
further subcase obtained by combining the two subclasses above.

The complete classification  for $N=4$ is relatively
straightforward, due to the facts that there always exists at least
one aligned direction, that all such aligned directions are
discrete, normalization reduces the possible number of subclasses
(leading to unique subcases) and since reducibility is not an
issue because a Weyl tensor of a manifold of
dimension 3 or less must necessarily vanish. The present classification reduces to the classical
Petrov classification in 4 dimensions \cite{Algclass}.

In most applications \cite{RT,soliton,brane}
the Weyl
classification is relatively simple and the details of the more complete
classification are not necessary. It would be useful
to be able to find a more practical way of determining the
Weyl type, such as for example employing certain scalar
higher dimensional invariants. We also note that it may be more practical in some
situations to classify the Riemann tensor
since it is not
considerably more difficult to classify the Riemann tensor rather
than the Weyl tensor in higher dimensions \cite{Algclass}.

\subsection{Petrov classification in four dimensions}

The Petrov classification \cite{petrov} (see also \cite{penrind,Penrose}), 
invariantly determines the algebraic type of the
Weyl tensor at a given point of a 4-dimensional spacetime. A
spacetime is said to be of a given Petrov type if the type is the same at all
points. There are several equivalent methods of obtaining the
classification. The most widely used method is the two component
spinor approach, developed by Penrose {\cite{penrind,Penrose}}, in which
the Weyl tensor is represented by a totally
symmetric spinor and consequently can be decomposed in terms of
four principal spinors. Petrov types now correspond to various
multiplicities of principal spinors. Null vectors corresponding to
various principal spinors are  called {\it principal null vectors}
and the corresponding directions are called {\it principal null directions}
(PNDs). Thus there are at most four distinct PNDs at each point of
a spacetime.

In four dimensions an equivalent but distinct method leading to
the Petrov classification in terms of alignment type was given
in \cite{Algclass}. In four dimensions, in the
standard complex null tetrad (${\bl}$, ${\vn}$, $\btm$,  $\cbm$)
\cite{penrind,PravProc,kramer}, the Weyl tensor has five complex components.
Components of the Weyl tensor have
different boost weights (as defined by eqns.
(\ref{eq:boost}) and (\ref{eq:boostxform})). When $\bl$ coincides with the PND, then
the $+2$ boost weight terms vanish.
For the algebraically special types {\bf I}, {\bf II}, {\bf III}, {\bf N}, components with boost
weight greater or equal to
2,1,0,-1, respectively, can be
transformed away (see Table 2). Type {\bf D} is a special subcase of type {\bf II} in which
all components with non-zero boost weight can be transformed away.

In this approach the top alignment equations can be rewritten in terms of
an associated fourth degree polynomial of one
complex variable,
$\Psi(z_+)=0$.
It follows that the equations for
alignment order $q$ have a solution if and only if $\Psi(z_+)$
possesses a root of multiplicity $q+1$ or more.  In this way
the usual Petrov classification, which counts the root
multiplicities of the polynomial $\Psi(z_+)$, is recovered.
Of course, the two versions are equivalent
in four dimensions \cite{Algclass}.
However, it is this alignment type version of the Petrov classification that we have
generalized to higher dimensions.

\subsubsection{Classification of the Weyl tensor in higher dimensions:}

The WANDS are a natural generalization of the PNDs in higher
dimensions, and the classification of the Weyl tensor is
based  on the existence of WANDs of various orders of alignment.
We again note that once we fix $\bl$ as a WAND with  maximal order
of alignment, we can search for ${\vn}$ with maximal order of
alignment subject to the constraint ${\vn} \cdot \bl=1$   and
similarly define {\it secondary alignment type}. Alignment type is
a pair consisting of primary and secondary alignment types.
The possible alignment types are summarized in Table 2 (taken from
\cite{Algclass,PP}); the link with the four
dimensional Petrov classification is emphasised.
In five dimensions and higher, the generic
Weyl tensor does not have any aligned directions. Thus, unlike
the case in four dimensions, type {\bf I} tensors are an algebraically special category.

\begin{table}[h]
\begin{center}
\begin{tabular}{|cc|cc|}
\hline
\ \ D$>$4 dimensions & & 4 dimensions \\
\hline
Weyl type & alignment type& Petrov type & root multiplicities  \\
\hline
{\bf G}     & {\bf G} &
    \\
{\bf I}     & (1)   &  \\
{\bf I}$_{i}$ & (1,1) & {\bf I} & (1,1,1,1)\\
{\bf II}    & (2)  &  \\
{\bf II}$_{i}$ & (2,1) & {\bf II} & (2,1,1)\\
{\bf D} & (2,2) & {\bf D} & (2,2)\\
{\bf III} & (3) &  \\
{\bf III}$_{i}$ & (3,1) & {\bf III} & (3,1) \\
{\bf N} & (4) & {\bf N} & (4)\\
\hline
\end{tabular}
\caption{Algebraic classification of the Weyl tensor in four and
higher dimensions. Note that in four dimensions alignment types
(1), (2) and (3) are necessarily equivalent to the types (1,1),
(2,1) and (3,1), respectively, and since there is always at least
one PND type {\bf G} does not exist.}
\end{center}
\end{table}

\subsection{Methods of Classification}

There are essentially three methods currently available
to determine the Weyl type. In a straightforward approach the
alignment equations are studied, which are $\frac{1}{2}N(N-3)$
degree-4 polynomial equations in $(N-2)$ variables (and are
generally overdetermined and hence have no solutions for $N>4$),
to determine if there exist non-trivial solutions. A second
method, in which the necessary conditions are investigated \cite{PP}, is
more practical (and results in studying essentially the same
equations but perhaps in a more organized form and, in the case of type {\bf N}
for example, reduces to the study of linear equations); this approach is followed
in classifying the black ring solutions (see below).
Finally, in many applications which are simple generalizations of
4-dimensional solutions in which the preferred 4-dimensional null frame is explicitly
known, the 5-dimensional null frame can be determined (guessed) directly.
Some examples of this latter approach will be presented below. It is clear that these
current methods for finding WANDs are very cumbersome. It is
consequently important to derive a more practical method for
determining Weyl type; for example, by utilizing invariants as in
the case of 4 dimensions \cite{kramer}.

A {\em curvature invariant of order $n$} is a scalar obtained by
contraction from a polynomial in the Riemann tensor and its
covariant derivatives up to the order $n$. The Kretschmann scalar,
$R_{abcd} R^{abcd}$,    is an example of a zeroth order invariant.
In Riemannian geometry the vanishing of the Kretschmann scalar invariant implies
flat space. Invariants have been extensively used in the study of $VSI$ and $CSI$ 
spacetimes.

In 4 dimensions, demanding that the zeroth order quadratic and cubic Weyl invariants $I$ 
and $J$ vanish
($I=J=0$) implies that the Petrov type is {\bf III}, {\bf N}, or {\bf  O}.
In higher dimensions,  all of the zeroth order invariants vanish in type
{\bf III}, {\bf N} and {\bf O} spacetimes. It would be particularly useful to find
necessary conditions in terms of simple invariants for a type  {\bf I} or
type  {\bf II} spacetime. 

\subsubsection{Necessary conditions for WANDs:}

The Weyl classification is based on the existence of preferred
WANDs. The necessary conditions for
various classes, which can significantly simplify the search for
WANDs, are as follows \cite{PP}:

\BEA
\begin{tabular}{rcl}
$\ell^b \ell^c \ell_{[e}C_{a]bc[d}\ell_{f]}=0$ & $\Longleftarrow  $ & $\ell$ is WAND,
at most primary type {\bf I};\\
$\ell^b \ell^c C_{abc[d}\ell_{e]}=0$& $\Longleftarrow  $& $\ell$ is WAND,
at most primary  type {\bf II};\\
$ \ell^c C_{abc[d}\ell_{e]}=0$ & $\Longleftarrow  $ & $\ell$ is WAND,
at most primary type {\bf III};\\
$\ell^c C_{abcd}=0$ & $\Longleftarrow  $ & $\ell$ is WAND, at most primary type
{\bf N}.\\
\end{tabular}
\label{HDnecessity}
\EEA

In higher dimensions ($N>4$) the conditions given on the left-hand-side of
(\ref{HDnecessity}) are only necessary (but not necessarily sufficient)
conditions for the statements on the right-hand-side of
(\ref{HDnecessity}) \cite{PP}.
For type {\bf I}, equivalence holds in arbitrary dimension (see below), but this is
not so for more special types. (In four dimensions, all of these
relationships are equivalences).
For example, a spacetime satisfying   $\ell^c C_{abcd}=0$ can be
of type {\bf II} \cite{PP}; such a  spacetime has in principle
a non-vanishing curvature invariant $C_{abcd} C^{abcd}$ and for
type {\bf N} and {\bf III} spacetimes in arbitrary dimension all polynomial
curvature invariants constructed from components of the Weyl
tensor vanish.

There are $\frac{1}{2} N(N-3)$ independent scalars of maximal boost weight 2
for a Weyl tensor. The top alignment
equations, $C_{0i0j}=0$,
are a system of $\frac{1}{2} N(N-3)$, fourth order
equations in $N-2$ variables.
In \cite{Algclass} it was shown that, in any dimension $N$, a null vector $\vk$
satisfies the first equation of (\ref{HDnecessity}); i.e.,
\begin{equation}
  \label{eq:pndeq}
 k^b k_{[e} C_{a]bc[d}  k_{f]}k^c=0,\quad k^a k_a=0,
\end{equation}
if and only if $\vk$ is aligned with $C_{abcd}$ (i.e., the above system of equations is
equivalent to the Weyl alignment equations).
In 4 dimensions, these equations define the PNDs of the Weyl
tensor \cite{Penrose,kramer}.

\subsection{Classification of spacetimes}

Many higher dimensional spacetimes are now known, including a
number of vacuum solutions of the Einstein field equations which
can represent higher dimensional black holes. These N-dimensional
black holes are of physical interest, particularly in view of the
development of string theory. Let us review some of the higher
dimensional spacetimes that have been classified algebraically
(many of these results are taken from \cite{5Dclass}); the results
are summarized in Table 3.

Higher dimensional generalizations of the Schwarzschild solution,
the Schwarzschild-Tangherlini (ST) solutions \cite{tang}, which are
spherically symmetry on spacelike $(N-2)$-surfaces, are of
algebraic Weyl type {\bf  D}. In 5 dimensions this solution is the unique
asymptotically flat, static black hole solution with
spherical symmetry on spacelike 3-surfaces. Higher dimensional
generalizations of Reissner-Nordstrom black holes are also of type
{\bf D}.

A class of 5-dimensional Kaluza-Klein vacuum soliton solutions \cite{soliton},
the Sorkin-Gross-Perry-Davidson-Owen soliton (GP), are
also of physical interest. The non-black hole
solutions (i.e., all solutions except the 5-dimensional generalized
Schwarzschild (S$^*$) solution) are of type {\bf I} \cite{5Dclass}.
There is a special case, GP$_s$, which is of type {\bf D}.

The Myers-Perry solution (MP) in five and higher dimensions
\cite{Myers}, a direct generalization of the 4-dimensional aymptotically
flat, rotating Kerr black hole solution, is also of type {\bf D}. A
class of higher dimensional Kerr-(anti) de Sitter (K(A)S)
solutions, which are given in N-dimensions and have $(N-1)/2$
independent rotation parameters, have been given in Kerr-Schild
form \cite{page}. These rotating black hole solutions with a
non-zero cosmological constant reduce to the 5-dimensional solution of
\cite{hht} and the Kerr-de Sitter spacetime in 4 dimensions, and the
Myers-Perry solution in the absence of a cosmological constant.
The K(A)S spacetimes were shown to be of type {\bf  D} \cite{5Dclass}. Recently, it has been
shown that the family  of N-dimensional rotating black
holes with a cosmological constant and a NUT parameter (LNUT) are of type
{\bf D} \cite{DNUT}.

In all of the above examples a canonical null frame was found
\cite{5Dclass}, in which the Weyl basis components had the
appropriate form. In the following example the necessary
conditions (\ref{HDnecessity}) were studied directly to classify
the spacetimes \cite{PP}.

Non-rotating uncharged black string  Randall-Sundrum braneworlds
were first discussed in \cite{chamblim}. The rotating ``black ring''
solutions are vacuum, asymptotically flat,
stationary black hole solutions of topology $S^ 1 \times S^2$
\cite{ER-PRL}. These solution have subsequently been generalized
to the non-supersymmetric black ring solutions of minimal
supergravity in \cite{EEF}.

The necessary conditions
(\ref{HDnecessity}) can be used to classify the black ring
solution (BR). The method is to solve the necessity conditions and then
check that these solutions do indeed represent WANDs by
calculating the components of the Weyl tensor in an appropriate
frame \cite{PP}. In order to solve the first equation in
(\ref{HDnecessity}), $\ell^a$ is denoted by $(\alpha, \beta,
\gamma, \delta, \epsilon)$. A set of fourth order polynomial
equations in $\alpha \dots \epsilon$ is then obtained. An
additional second order equation follows from $\ell_a \ell^a=0$.
From a direct analysis of these equations, it was shown
that the black ring solution is algebraically special
and of type {\bf I}$_i$ \cite{PP}.
On the horizon, a transformation leads to a metric regular on the
horizon so that there is second solution to (\ref{HDnecessity}).
It can be checked that the boost order of the Weyl tensor in the
frame with this solution is zero,  and thus the black ring is of
type {\bf II} on the horizon (BR$_H$). By an appropriate choice of parameters we
obtain the Myers-Perry metric (MP) \cite{Myers} with a single rotation
parameter. It turns out that the second equation in
(\ref{HDnecessity}) then admits two independent solutions,
and the spacetime is thus of type {\bf D}. 

There are also supersymmetric rotating black holes that exist in
five dimensions. There is the extremal charged rotating BMPV black
hole of \cite{BMPV} in minimal supergravity, with a horizon of
spherical topology (see also  \cite{BMPV}). The BMPV metric is of
Weyl type {\bf I}$_i$. Recently, the static charged black
ring (CBR) \cite{Dida} has been shown to be of type {\bf G}
\cite{Dort}; this is the only explicitly known exact solution that
is algebraically general.

There are many other higher dimensional spacetimes of interest. In
arbitrary dimensions it is known that the Weyl tensor of a
spherically symmetric and static spacetime is "boost invariant",
and thus of type {\bf  D} \cite{HR}. Indeed, recently it has been shown
that all static  spacetimes (STAT) (and a class of stationary
speactimes) in dimensions $N > 4$ are necessarily of Weyl types {\bf G},
{\bf I}$_i$, {\bf  D} or {\bf O}, and that spherically symmetric spacetimes  (SS) are
of type {\bf  D} or {\bf O} \cite{PraPraOrt07}.

$VSI$ spacetimes are $N$-dimensional Lorentzian spacetimes in which
all curvature invariants of all orders vanish  \cite{Higher,CFHP}.
Higher-dimensional $VSI$ spacetimes are of Weyl and Ricci types {\bf N}, 
{\bf III} or {\bf O} \cite{Higher}. The Ricci type {\bf N} $VSI$ spacetimes include the
higher-dimensional Weyl type {\bf N} (generalized) pp-wave spacetimes (PP). A class
of higher dimensional relativistic gyratons (RG) \cite{frolov}, which are
vacuum solutions of the Einstein equations of the generalized
Kundt class (representing a beam pulse of spinning radiation), are
of type {\bf III}. A class of higher dimensional Robinson-Trautman  spacetimes (RT)
are of algebraic type {\bf  D}  \cite{Podolsky}.

In general, the Weyl and Ricci types of non-locally homogeneous $CSI$ spacetimes, in
which all curvature invariants of all orders are constant
\cite{CSI}, is {\bf II} \cite{CFH}. There are a number of special
examples of  $CSI$ spacetimes;  $AdS \times S$ spacetimes (AdSXS),
such as for example $AdS_5
\times S^5$ \cite{Freund}, are of Weyl type {\bf  D} (or {\bf O}) \cite{CFH},
generalizations of $AdS \times S$ based on different $VSI$
seeds (AdSXG) are of Weyl type {\bf III} if the
sectional curvatures are of equal magnitude and opposite sign
\cite{CFH}, and the
higher-dimensional $AdS$ gyratons (AdSG) \cite{frolov} are of Weyl type
{\bf III}.

Supergravity solutions with a covariantly constant null vector (CCNV) are
of interest in the study of supersymmetry \cite{TOrtin}. These CCNV
spacetimes are
a product manifold with a CCNV-$VSI$ Lorentzian piece of Ricci and
Weyl type {\bf III} and a locally homogeneous transverse Riemannian
space of Ricci and Weyl type {\bf II} \cite{CFHP}.

\newpage

\begin{table}
{\small
\begin{center}
\begin{tabular}{|l|l|l|l|c|c|c|} \hline
&&&&&&\\
{\bf Name} & {\bf Ref} & {\bf Comments} & {\bf Type}
& {\bf Dim} & {\bf Special cases} & {\bf Type}\\
&&&&&&\\ \hline
&&&&&&\\
ST & \cite{tang} & vacuum BH  & {\bf D} & N &&\\
&& $R^2 \times S^3$ &&&&\\
&&&&&&\\
SS & \cite{PraPraOrt07} & Sph. Sym. & {\bf D}/{\bf O} & N &&\\
&&&&&&\\
STAT & \cite{PraPraOrt07} & Static & {\bf G}/{\bf I}$_{i}$/{\bf D} & N &&\\
&&&&&& \\ \hline
&&&&&&\\
GP & \cite{soliton} & vacuum soliton   & {\bf I} & 5 & &\\
&& $R \times R^2 \times S^2$ &  & & & \\
&& &  & & S$^*$ & {\bf D}\\
&&&&&&\\
K(A)S & \cite{page} & Rotating BH & {\bf D} & N &   & \\
&& $\Lambda \neq 0$ & & &  &\\
& & & & &  & \\
& \cite{Myers}& & & N & MP & {\bf D}\\
LNUT& \cite{DNUT}& & {\bf D} & N && \\
&&&&&&\\ \hline
&&&&&&\\
BR & \cite{ER-PRL} & Rotating BR & {\bf I}$_i$ & 5 & & \\
&& $R \times R^2 \times S^2$ & & &  &\\
&&&&&&\\
& \cite{PP} &  &  &  & BR$_H$ & {\bf II}\\
& \cite{Myers}&  & & 5 & MP & {\bf D}\\
&&&&&&\\
BMPV & \cite{BMPV} & Supersymmetric & {\bf I}$_i$ & 5 &&\\
&&&&&&\\
CBR & \cite{Dort} & Charged & {\bf G} & 5 &&\\
&&&&&&\\
\hline
&&&&&&\\
VSI & \cite{Higher} & Non BH & {\bf N}/{\bf III} & N &&\\
&&&&&&\\
&  &  &  &  & PP & {\bf N}\\
& \cite{frolov} &  &  &  & RG & {\bf III}\\
&&&&&&\\
CSI &  \cite{CFH} &  & {\bf II} & N &&\\
& &  &  &  & AdSXG & {\bf III}\\
& &  &  &  & AdSXS & {\bf D}/{\bf O}\\
& \cite{frolov} &  &  &  & AdSG & {\bf III}\\
&&&&&&\\
RT &  \cite{Podolsky} &  & {\bf D} & N &&\\
CCNV &  \cite{CFH} &  & {\bf II} & N &&\\
&&&&&&\\ \hline
\end{tabular}
\end{center}}
\caption{The solution ("Name") is identified by the acronym given in
the text. In the comments, some features of the solution are presented;
i.e., whether it is a black hole (BH) and whether or not it is
rotating, whether there is a non-zero cosmological constant
$\Lambda$, its topology etc. In the "Dim"
column, it is indicated whether there are arbitrary dimensional
generalizations of the solution ($N$) or the solution is
for a specific dimension (e.g., $N=5$).}
\end{table}

\newpage

\section{Higher Dimensional Frame Formalism}

A higher dimensional frame formalism has been developed
\cite{Bianchi,ricci,Higher} in order to completely generalize the
Newman--Penrose (NP) formalism \cite{NP,kramer,penrind} for any
$N>4$. The main focus of the review of the classification of the Weyl tensor
so far has been algebraic.
However, there is a rich interplay between algebraic type and
differential identities. Indeed, in higher dimensions it is possible to
use the Bianchi and Ricci equations to
construct algebraically special solutions of Einstein's field
equations, at least for the simplest algebraically special
spacetimes.

The Riemann tensor is a 4-rank tensor  with the symmetries \BE
R_{abcd}=\frac{1}{2} (R_{[ab] [cd]} + R_{[cd] [ab]}),
\label{Rsym1} \EE such that $R_{a\{bcd\}} \equiv
R_{abcd} + R_{acdb} + R_{adbc}=0$.  We can also
decompose the Riemann tensor in its frame components and sort them
by their boost weights in the same way as is done for the Weyl
tensor \cite{Algclass}.
The effects of null rotations, boosts and spins on
the components of the Riemann tensor are given in \cite{Algclass,Bianchi}.

Constraints on the Riemann tensor and the Ricci
rotation coefficients are obtained from the commutation relations
\cite{Higher}, the Bianchi identities (or more percisely the
Bianchi equations) \cite{Bianchi} \BE R_{ab\{cd;e\}} \equiv
R_{abcd;e}+R_{abde;c}+R_{abec;d} =  0, \label{Bia} \EE and the
Ricci identities \BDM V_{a;bc}=V_{a;cb}+R^s_{\ abc} V_s , \EDM
where ${\mbox{\boldmath{$V$}}}$ is an arbitrary vector
\cite{ricci}.

In 4 dimensions, for algebraically special vacuum spacetimes some of the
tetrad components of the Bianchi identities in the Newman-Penrose
formalism \cite{NP} lead to simple algebraic equations (i.e.,
equations with no derivatives). In higher dimensions these
algebraic equations are much more complex and the number of
independent equations, as well as the number of unknowns, depend
on the dimension of the spacetime. In addition, in 4 dimensions it is
possible to use the Bianchi and Ricci equations to construct many
algebraically special solutions of Einstein's field equations. It
is conceivable that a similar thing is possible in higher
dimensions, at least for the simplest algebraically special
spacetimes. Indeed, the vast majority of today's known higher
dimensional exact solutions are simple generalizations of 4-dimensional
solutions.

\subsection{Preliminaries}
Let us denote the components of the covariant derivatives of the frame
vectors $\bl,{\vn}, \bm^{(i)}$ by $L_{ab},N_{ab}$ and $\Mi_{ab}$
(the Ricci rotation coefficients), respectively:
\BDM
\fl
\ell_{a;b}=L_{cd} m^{(c)}_a m^{(d)}_{b} \ , \ \ n_{a;b}=N_{cd} m^{(c)}_a m^{(d)}_{b} \ , \ \
m^{(i)}_{a;b}=\Mi_{cd} m^{(c)}_a  m^{(d)}_{b}  \ 
\EDM
(where we follow the notation of \cite{Bianchi}).
Since the norm of  all frame vectors is constant,  it follows that
\BDM
L_{0a}=N_{1a}=\Mi_{ia}=0.
\EDM
Also from the fact that all scalar products of the frame vectors are constant, we get
\BE
\!\!\!\!\!\!\!\!\!\!\!\!\!\!\!\!\!\!
N_{0a}+L_{1a}=0, \quad \Mi_{0a} + L_{ia} = 0, \quad \Mi_{1a}+N_{ia}=0,
\quad \Mi_{ja}+\Mj_{ia}=0. \label{const-scalar-prod}
\EE
Therefore, there are
$\frac{1}{2}N^2(N-1)$ independent rotation coefficients. (In $N=4$ dimensions the
$L_{ab}$, $N_{ab}$ and $\Mi_{ab}$ are equivalent to the 12 complex NP spin
coefficients.)

The covariant derivatives of the frame vectors can then be expressed as
\cite{Bianchi}:
\BEA
\fl
\ell_{a ; b} &=& L_{11} \ell_a \ell_b + L_{10} \ell_a n_b + L_{1i} \ell_a m^{(i)}_{\, b}  +
L_{i1} m^{(i)}_a \ell_b   + L_{i0} m^{(i)}_{\, a} n_{b} + L_{ij} m^{(i)}_{\, a} m^{(j)}_{\, b}  , \label{dl} \\
\fl
n_{a ; b  } &=&\! -\! L_{11} n_a \ell_b -\! L_{10} n_{a } n_{b }-\! L_{1i} n_a m^{(i)}_{\, b}  +
N_{i1} m^{(i)}_{\, a} \ell_b  + N_{i0} m^{(i)}_{\, a} n_b  + N_{ij} m^{(i)}_{\, a} m^{(j)}_{\, b}  , \label{dn} \\
\fl
m^{(i)}_{a ; b } &=&\! -\! N_{i1} \ell_a \ell_b -N_{i0} \ell_a n_b -{L}_{i1} n_a \ell_b
-L_{i0} n_a n_b -N_{ij} \ell_a m^{(j)}_{\, b}   \nonumber \\
\fl
&&\! +\! {\Mi}_{j1} m^{(j)}_{\, a} \ell_b  -L_{ij} n_a m^{(j)}_{\, b}
+ {\Mi}_{j0} m^{(j)}_{\, a} n_b + {\Mi}_{kl} m^{(k)}_{\, a} m^{(l)}_{\, b} . \label{dm}
\EEA
The transformation properties of the rotation coefficients under
Lorentz transformations (null rotations, spins and boosts) are given in \cite{ricci}.

The directional derivatives $D$, $\T$, and
$\delta_i$ can then be introduced:  \BE D \equiv \ell^a \nabla_a, \ \ \ \bigtriangleup
\equiv n^a \nabla_a, \ \ \ \delta_i \equiv \msup{i}{a} \nabla_a ,
\quad \nabla_a = n_a D + \ell_a \T + \msub{i}{a} \delta_i.
\label{covder} \EE
The commutators then have the form \cite{Higher} \BEA
\T D - D \T &=& L_{11} D + L_{10} \T + L_{i1} \delta_i - N_{i0} \delta_i, \\
\delta_i D - D \delta_i &=& (L_{1i} + N_{i0}) D + L_{i0}
\T + (L_{ji}-
{{M}\hspace*{-.12in}\raisebox{.11in}{\em i}\hspace*{.05in}_{j0}})
\delta_j, \\
\delta_i \T - \T \delta_i  &=& N_{i1} D + (L_{i1}-L_{1i}) \T +
(N_{ji}-
{{M}\hspace*{-.12in}\raisebox{.11in}{\em i}\hspace*{.05in}_{j1}})
\delta_j, \\
\delta_i \delta_j - \delta_j \delta_i &=& (N_{ij}-N_{ji}) D + (L_{ij}-L_{ji})\T
 + (
 {{M}\hspace*{-.12in}\raisebox{.11in}{\em j}\hspace*{.05in}_{ki}}-
 {{M}\hspace*{-.12in}\raisebox{.11in}{\em i}\hspace*{.05in}_{kj}})
 \delta_k.
\EEA

\subsubsection{The frame components of the Bianchi equations and the Ricci
identities:}

Contractions of the Bianchi equations and the Ricci identities
with various  combinations of the frame vectors lead to a set of first order
differential equations presented in \cite{Bianchi,ricci}. For algebraically general
spacetimes these equations are quite complicated.  However, for
algebraically special cases they are much simpler.
For $N=4$ these equations are equivalent to the standard Bianchi and Ricci identities arising
in the NP formalism \cite{kramer}.

From a study of the algebraic  Bianchi
equations, it was shown \cite{Bianchi} that in vacuum type {\bf III} and
{\bf N} spacetimes of arbitrary dimension the multiple principal null
direction (PND) is geodesic. It was also shown that for type {\bf N} and
{\bf III} vacuum spacetimes the expansion and twist matrices $\mS$ and
$\mA$ (defined in the next subsection) have very specific properties (described in detail in
\cite{Bianchi}; however, note that the shear does not vanish for
$N>4$). Conversely, it was shown that for a vacuum spacetime the
properties of the $\mS$ and $\mA$ matrices mentioned above imply that
the spacetime is algebraically special. The complete analysis,
including all possible degenerate cases, was presented for
5-dimensional spacetimes in \cite{Bianchi}.

The optical properties of the gravitational field in higher
dimensions were further explored in \cite{ricci}. It was shown
that it follows immediately from the Ricci identities and the
Sachs equations in arbitrary dimensions \cite{ricci} that (under
the assumption $R_{00}=0=R_{0i}$ on the matter fields) $N\ge 4$
dimensional Kundt spacetimes are of type {\bf II} (or more special), and
for odd $N$ a twisting geodetic WAND must also be shearing (in
contrast to the case ${N=4}$).

\subsection{Null geodesic congruences}

The congruence corresponding to $\bl$ is geodesic if \BDM \ell_{a
; b } \ell^{b} \propto \ell_{a } \   \Leftrightarrow \  L_{i0} =
0. \EDM It is always possible to rescale $\bl$ in such a way that $\ell_{a ; b } \ell^{b} = 0$ and
thus also $L_{10}=0$ (i.e., $\bl$ is {\em affinely parametrized}).
Using this parametrization, the covariant derivative of the vector
$\bl$ is then \BE \ell_{a ; b } = L_{11} \ell_a \ell_b +L_{1i}
\ell_a m^{(i)}_b  + L_{i1} m^{(i)}_a \ell_b  + L_{ij} m^{(i)}_{a}
m^{(j)}_{b}. \label{dlgeod} \EE
We can decompose ${\mL}$ into its
symmetric and antisymmetric parts, ${\mS}$ and ${\mA}$, \BDM
L_{ij}=S_{ij} + A_{ij}, \quad S_{ij} = S_{ji} , \ A_{ij}= -A_{ji}
, \EDM where \BDM S_{ij}=\ell_{(a ; b)} m^{(i)a} m^{(j)b} \ , \ \
\ A_{ij}=\ell_{[a ; b]} m^{(i)a} m^{(j)b}. \EDM

We can then define
the expansion  (trace) $\theta$ and the shear (tracefree
symmetric) matrix $\sigma_{ij}$ as follows:
\BE
\theta \equiv \frac{1}{N-2} \ell^{a}_{\ ; a}\ ,
\sigma_{ij} \equiv \left( \ell_{(a ; b)} - \theta \delta_{kl}
m^{(k)}_a m^{(l)}_b \right) m^{(i)a} m^{(j)b}
=S_{ij}- \theta  \delta_{ij} \label{shearmatrix} \EE
(i.e., $L_{ij}=\sigma_{ij}+\theta\delta_{ij}+A_{ij}$). For
simplicity, let us call ${\mbox{\boldmath $A$}}$ the {\it twist
matrix} and ${\mbox{\boldmath $S$}}$  the {\it expansion matrix},
although ${\mbox{\boldmath $S$}}$ contains information about both expansion and shear.
In addition to
$\theta$, we can construct other (optical) scalar quantities from
$\ell_{a ; b}$ which are invariant under null and spatial
rotations with fixed $\bl$; e.g., the shear and twist scalars
given by the traces
$\sigma^2\equiv\sigma^2_{ii}=\sigma_{ij}\sigma_{ji}$ and
$\omega^2\equiv-A^2_{ii}=-A_{ij}A_{ji}$ (note that
$\sigma_{ij}=0\Leftrightarrow\sigma^2=0$ and
$A_{ij}=0\Leftrightarrow\omega^2=0$).

We note that it is always possible to choose ${\vn}$ and $\bm{_i} $
to be {\em parallely propagated} along the geodesic null
congruence $\bl$; i.e., we can can set $\M{i}{j}{0}=0$ and
$N_{i0}=0$ by performing an appropriate spin transformation and
then a null rotation. This, for example, simplifies the Ricci
identities considerably and may be convenient for certain
calculations.

Finally, for a geodetic null congruence, splitting the equations
into their tracefree symmetric, trace and antisymmetric parts, we
obtain the set of $N$-dimensional {\em Sachs equations} for
$D\sigma_{ij}$, $D\theta$ and $ DA_{ij}$ \cite{ricci}.

\newpage

\subsection{Applications}

There are many applications of the Weyl classification scheme.
Higher dimensional $VSI$ spacetimes are necessarily of Weyl type
{\bf III}, {\bf N} or {\bf O}, and higher dimensional $CSI$ spacetimes
that are not locally homogeneous are
at most of type {\bf II}; these results have been extremely useful in
the study of $VSI$ and $CSI$ spacetimes (see the next section). In addition,
in many applications the higher
dimensional frame formalism has proven to be crucial.

\subsubsection{Goldberg-Sachs theorem:}

In 4 dimensions an important connection between geometric optics and the
algebraic structure of the Weyl tensor is provided by the
Goldberg-Sachs theorem \cite{NP,Sachs61,Goldberg}, which states that a
vacuum metric is algebraically special if and only if it contains
a  geodesic shearfree null congruence ~\cite{kramer,penrind}. This
theorem has played a fundemental role in the construction and
classification of exact solutions of Einstein's equations in four
dimensions. In addition, given an NP tetrad and assuming that this result applies to
both ${\bl}$ and $\bf n$, then the Weyl tensor is of type {\bf  D} with
repeated PNDs ${\bl}$ and $\bf n$ that are geodesic and shear-free. There are generalizations of the Goldberg-Sachs
theorem to non-vacuum spacetimes in 4 dimensions \cite{kramer}.

A higher-dimensional version of the Goldberg-Sachs
theorem is of considerable interest.
Unfortunately, the Goldberg-Sachs
theorem does not have a straightforward generalization for higher
dimensions  \cite{Frolov,Bianchi},
and any higher
dimensional version of the Goldberg-Sachs theorem will necessarily
be more complicated.
However, a partial extension of the Goldberg-Sachs
theorem to $N>4$ has been obtained
by decomposing the first covariant derivatives of
$\bl$ as in equation (\ref{dlgeod}) (where ${\bl}$ is assumed to be geodesic and affinely
parametrized), utilysing the
algebraic type of the Weyl tensor and studying the consequences
following from the Bianchi and Ricci identities \cite{Bianchi}.

It has been found that in vacuum, where the Riemann tensor is equal to the Weyl tensor,
the Bianchi identities for type {\bf N} spacetimes imply that in
arbitrary dimension the congruence corresponding to the WAND is
geodesic \cite{Bianchi}. It also follows that in higher dimensions
for vacuum type {\bf N} spacetimes with non-vanishing expansion the
matrices $S_{ij}$, $A_{ij}$ and the Weyl tensor have certain
specific forms and that the shear does not vanish.
Similar results apply for the multiple WAND of a type
{\bf III} vacuum spacetime \cite{Bianchi}.
Therefore, in $N>4$ vacuum spacetimes of type {\bf III} or {\bf N}, a
multiple WAND with expansion necessarily also has
non-zero shear \cite{Bianchi}. This constitutes a
non-zero shear `violation' of the 4-dimensional Goldberg-Sachs theorem in higher
dimensions. In addition, there exist explicit examples  of special vacuum type {\bf  D} spacetimes
in $N \geq 7$ admitting non-geodetic multiple WANDs
\cite{PraPraOrt07}; this shows that there are `violations' of the
geodetic part of the Goldberg-Sachs theorem in higher dimensions
(in addition to the non-zero shear `violations').

For example, the WANDs of 5-dimensional vacuum rotating
black hole spacetimes (which are algebraically special and of type {\bf  D}) are geodetic but shearing
\cite{Frolov,Bianchi}.
This result constitutes an explicit counterexample to a complete higher
dimensional extension of the Goldberg-Sachs theorem; in particular, this counterexample includes
the higher dimensional Weyl type {\bf  D} Kerr shearing vacuum metric
and the $N=5$ type {\bf  D} Myers-Perry twisting and shearing
 vacuum black hole spacetime \cite{Frolov,Bianchi}.
In addition, we note that a non-zero
$\sigma_{ij}$ does not imply a non-zero $A_{ij}$;  for example, static black strings
and static black branes are type {\bf  D}
spacetimes with expanding, shearing but non-twisting geodetic
multiple WANDs \cite{PraPraOrt07}. Note that the shear in arbitrary dimensional Myers-Perry
spacetimes was studied in the Appendix of \cite{PraPraOrt07}.

\subsubsection{The type-{\bf  D} conjecture:}

Let us first summarize the 4-dimensional static and stationary black hole solutions
(with topology $S^2$); there is the Schwarszchild solution, the
more general Kerr-Newman solution, the non-vacuum
Reissner-Nordstrom spacetime, and the non-asymptotically flat
vacuum solutions such as the Schwarszchild-de Sitter spacetime. All of
these solutions are highly symmetric and are of Weyl (Petrov)
type {\bf  D} and hence algebraically special \cite{kramer}.
All of the known higher dimensional black
holes also have a great deal of symmetry \cite{reall}. It is
anticipated that this will again be reflected in their having
special algebraic properties. Indeed, all of the  higher
dimensional black holes classified so far are of algebraically
special (Weyl) type. In addition, as is the case in 4 dimensions
\cite{kramer},
in dimensions $N > 4$ it is known that the Weyl tensor of a
spherically symmetric spacetime is
of type {\bf  D} \cite{PraPraOrt07,HR}. This has led
to a conjecture that asserts that stationary higher dimensional
black holes, perhaps with the additional conditions of vacuum
and/or asymptotic flatness, are necessarily of Weyl type {\bf  D}
\cite{5Dclass,PP}.

This conjecture has received support recently from a study of local
(so that the results may be applied to surfaces of arbitrary
topology) non-expanding null surfaces (generic isolated horizons)
\cite{JLTP}. Assuming the usual energy inequalities, it was found
that the vanishing of the expansion of a null surface implies the
vanishing of the shear so that a covariant derivative is induced
on each non-expanding null surface. The induced degenerate metric
tensor, locally identified with a  metric tensor defined on the
$(N-2)$-dimensional tangent space, and the induced covariant
derivative, locally characterized by the rotation 2-form in the
vacuum case, constitute the geometry of a non-expanding null
surface. The remaining components of the surface covariant
derivative lead to constraints on the induced metric and the
rotation 2-form in the vacuum extremal isolated null surface case.
This leads to the condition that at the {\em non-expanding
horizon} the boost order of the null direction tangent to the
surface is at most $0$, so that the Weyl tensor is at most of type
{\bf II} \cite{JLTP} (where the {\it aligned} null vectors tangent to
the surface correspond to a double PND
of the Weyl tensor in the 4-dimensional case). For example, vacuum black
rings \cite{ER-PRL}, which are of type {\bf I}$_i$ in the stationary
region, are of Weyl type {\bf II} on the horizon \cite{PP}.

There has been some interesting recent work on hidden symmetries
that may be related to the type-{\bf D} conjecture \cite{hidden}.
Killing tensors are directly associated with conserved quantities for 
geodesic motion. More fundamental hidden symmetries are connected with 
antisymmetric  Killing-Yano tensors. Hidden symmetries
play an important role in the separability
of the Hamilton-Jacobi and Klein-Gordon equations. In 4 dimensions
the relations between hidden symmetries, the special algebraic type of
spacetimes, and the separation of variables for various field equations are well
known \cite{kramer}. It is of interest to study which of these properties carry 
over to spacetimes with more than four dimensions; that is, to 
study the relationship between the class of spacetimes with hidden symmetries
and their algebraic type in higher dimensions. The origin of 
hidden symmetries in higher dimensions is related to the existence of a 
so-called principal Killing-Yano tensor \cite{hidden}.
The most general Kerr-NUT-(A)dS black hole
spacetimes are of
Weyl type {\bf D} \cite{DNUT}. 
It has recently been
shown that, starting with the principal Killing-Yano tensor, 
it is possible to prove the complete integrability of geodesic motion
in these spacetimes and, moreover,  to
demonstrate this integrability  by explicitly separating the Hamilton-Jacobi and
Klein-Gordon equations \cite{hidden}.

\subsection{Peeling theorem in higher dimensions}

There is interest in the asymptotic ``peeling properties'' of the
Weyl tensor in higher dimensions. In 4-dimensional spacetimes it has
proven very useful to utilize the conformal structure of
space-time in studying asymptotic questions in general relativity, which include a
geometrical definition of asymptotic flatness and covariant
definitions of incoming and outgoing gravitational waves in vacuum
gravitational fields due to an isolated gravitating system \cite{NP}.
The ``peeling property'' in higher dimensions in the case of even
dimensions (and with some additional assumptions) was demonstrated
in \cite{peeling}, thereby providing a first step towards an
understanding of the general peeling behaviour of the Weyl tensor,
and the aymptotic structure at null infinity, in higher
dimensions. A more rigorous analysis of the peeling theorem in
higher dimensions would be
desirable.

Let us consider an $N$-dimensional spacetime ($M$, $g_{ab}$), $N$
even, that is weakly asymptotically
 simple at null
infinity \cite{holl}. The metric of an unphysical  manifold
(${\tilde M}$, ${\tilde g}_{ab}$) with boundary  $\Im $, is
related to the physical metric by a  conformal transformation,
${\tilde g}_{ab}=\Omega^2 g_{ab}$, where $\Omega=0$  and  ${\tilde
n}_a\equiv -\Omega_{;a}\not= 0$, null,  at   $\Im $. Let us make
the further natural assumption in higher dimensions that
components of the unphysical Weyl tensor with respect to the
unphysical tetrad are of order ${\cal O}(\Omega^{{ q}}) $ (with
$q\geq 1 $) in the neighbourhood of $\Im $ (in 4 dimensions  $q=1$, which
follows from Einstein's equations). Let
${\tilde \gamma}\subset ({\tilde M},\ {\tilde g}_{ab})$ be a null geodesic in the
 unphysical manifold
that has an affine parameter  ${\tilde r}\sim -\Omega$ near $\Im$
and a tangent vector ${\bl}$ and  let  ${ \gamma}\subset ({ M},\ {g}_{ab})$ be a corresponding null geodesic in the
physical manifold
with an affine parameter   $r\sim 1/ \Omega$ near $\Im $ and a
tangent vector $\tilde{\bl}$. We choose the frame in the physical
spacetime to be parallelly propagated along   ${\gamma}$   with
respect to ${g}_{ab}$ and the corresponding  frame $\tilde{\bl},
\tilde{\bf n}, \tilde{\bf m}^i$ in the unphysical spacetime to be related
by\BE {\tilde \ell}_a=\ell_a,\ \ \ {\tilde m}_a=\Omega m_a,\ \ \
{\tilde n}_a=\Omega^2n_a\ \\
\rightarrow {\tilde \ell}^a=\Omega^{-2}\ell^a,\ \ \
{\tilde m}^a=\Omega^{-1} m^a,\ \ \
{\tilde n}^a=n^a.\label{frameOmega}
\EE

The Weyl tensor in its frame components can then be sorted
according to their boost weight:\\
\BEA
  {{\tilde C}_{abc}}\ ^d&=&{\tilde g}^{de} {{\tilde C}_{abce}}
  ={\tilde g}^{de} [
      4 {\tilde C}_{0i0j}\, {\tilde n}^{}_{\{a}
      {\tilde m}^{(i)}_{\, b}  {\tilde n}^{}_{c}
      {\tilde m}^{(j)}_{\, e\: \}} + \cdots]
 \nonumber \\ &=&\Omega^{-2}{g}^{de}  \biggl[
      \Omega^{2+1+2+1} 4 {\tilde C}_{0i0j}\,
      {n}^{}_{\{a} {m}^{(i)}_{\, b}  {n}^{}_{c}  { m}^{(j)}_{\, e\: \}}+ \cdots\biggr]
      =C_{abc}\ ^d.\nonumber \EEA      Since all unphysical
components of the Weyl tensor ${\tilde C}_{1i1j}$, ${\tilde
C}_{1ijk}$, ${\tilde C}_{101i}$, ${\tilde C}_{ijkl}$, ${\tilde
C}_{0i1j}$, ${\tilde C}_{01ij}$, ${\tilde C}_{0101}$, ${\tilde
C}_{0ijk}$, ${\tilde C}_{010i}$, ${\tilde C}_{0i0j}$ are of order
${\cal O}(\Omega^q)$, each physical component is of order ${\cal
O}(\Omega^{b+2+q})$ (where $b$ denotes boost weight), and
 we obtain the peeling property \cite{peeling}
\BEAH
{ C}_{1i1j}&=&{\cal O}(\Omega^q),\\
{ C}_{1ijk},\ { C}_{101i}&=&{\cal O}(\Omega^{q+1}), \\
{ C}_{ijkl},\
{ C}_{0i1j},\
{ C}_{01ij},\
{ C}_{0101}&=&{\cal O}(\Omega^{q+2}), \  \\
{C}_{0ijk},\ { C}_{010i}&=&{\cal O}(\Omega^{q+3}), \ \\
{ C}_{0i0j}&=&{\cal O}(\Omega^{q+4}),
\EEAH
and thus
\BEA
  C_{abc}\ ^d &=& \Omega^q {C^{(N)}}_{abc}\ ^d + \Omega^{q+1} {C^{(III
)}}_{abc}\ ^d
+\Omega^{q+2} {C^{(II)}}_{abc}\ ^d+\Omega^{q+3} {C^{(I)}}_{abc}\ ^d\nonumber\\
&+& \Omega^{q+4}{C^{(G)}}_{abc}\ ^d +{\cal O}(\Omega^{q+5} ).
\EEA

Recently, this higher-dimensional formalism has been utilized to establish a formal 
analogue of the Weyl tensor peeling theorem in arbitrary dimensions in the
Penrose limit context \cite{blau}.

\newpage

\subsection{Further applications}

It is of interest to study the 5-dimensional case
and type {\bf D} spacetimes in more detail. In particular, we could
look for exact 5-dimensional solutions
that are generalizations of particular 4-dimensional
type {\bf D} solutions of special interest.

\subsubsection{Classification of the Weyl tensor in 5-dimensions:}

A spinorial formulation of 5-dimensional geometry was given in \cite{robspin} in order
to classify the Weyl tensor  in 5 dimensions (i.e., in an attempt to introduce
sufficiently many normalizations so that corresponding normal
forms become truely canonical).

Similar to the case of 4 dimensions, N=5 is special and can be equipped with a
compatible symplectic structure so that in 5 dimensions spinors can be used
to create a vector respresentation. Indeed, in 5 dimensions the (rank 4) Weyl
tensor can thus be represented by the Weyl spinor, which is
equivalent to a Weyl polynomial (homogeneous quartic polynomial in
3 variables). The classification is then realised by putting the
Weyl polynomial into a normal form; this amounts to determining
the aligned spinors or principle spin directions (PSDs) by solving
3 homogeneous quartic polynomials in 4 variables, and classifying
according to the nature of the solutions (i.e., the singularities of
the quartic, planar curves up to projective transformations) and hence
the multiplicities of the PSDs.

This 5-dimensional spinorial formulation is related to 
De Smet's polynomial
representation of the Weyl tensor \cite{DeSmet}. 
We recall that the De Smet classification is 
restricted  to static spacetimes and five dimensions (only). In addition, 
note that in five dimensions the De Smet classification \cite{DeSmet}
is not equivalent to our Weyl tensor
classification.  For example, the
simple 5-dimensional vacuum Robertson-Trautman solution is
algebraically special and of type {\bf III}
in our classification but
is algebraically general in De Smet's classification \cite{thesis}.

\subsubsection{Type {\bf  D} spacetimes:}

In general, the Weyl tensor has $\frac{1}{12}(N-2)(N-3)(N^2-3N+8)$ independent
components in type {\bf  D} spacetimes in $N$ dimensions.
Type {\bf  D} Einstein spaces were studied in
\cite{PraPraOrt07},  with an
emphasis on an investigation of the properties of WANDs using the Bianchi
identities (which become algebraic equations in this case). In particular, it was
shown that the multiple WAND in a vacuum type {\bf  D} spacetime is
geodetic in the `generic' case (as defined in \cite{PraPraOrt07})
in higher dimensions. However, special (i.e., not `generic')
vacuum type {\bf  D} spacetimes can admit non-geodetic multiple WANDs;
explicit examples in $N \geq 7$ were constructed in
\cite{PraPraOrt07}. Shear-free vacuum type {\bf  D} spacetimes  were also
studied.

Vacuum type {\bf  D} spacetimes have also been studied in the particular case of
five dimensions. In 5 dimensions, the type {\bf  D} Weyl tensor is fully determined
by a $3 \times 3$ matrix $\Phi_{ij}$. It is possible to further
algebraically classify the type {\bf  D} (or type {\bf II}) spacetimes in terms
of $\Phi_{ij}$ in this case \cite{PraPraOrt07,robspin}. The
relationship between shear-free spacetimes, twist-free spacetimes
and the properties of the $\Phi_{ij}$ in vacuum and Einstein
spaces were investigated in \cite{PraPraOrt07}; in particular, the multiple WANDs are
geodetic in 5-dimensional type {\bf  D} spacetimes.

\newpage

\section{VSI and CSI Spacetimes in Higher Dimensions}

Let us discuss the classes of  higher dimensional 
Lorentzian spacetimes with vanishing  scalar curvature
invariants ($VSI$ spacetimes) \cite{CFHP,Higher} and constant scalar
curvature invariants ($CSI$ spacetimes) \cite{CSI}. There are two
important applications of  $VSI$ and $CSI$ spacetimes. (i)
The~equivalence problem and the~classification of spacetimes; the
characterization of the $VSI$ and $CSI$ spacetimes may be a useful
first step toward addressing the~important question of when a
spacetime can be uniquely characterized by its curvature
invariants \cite{VSI}. (ii) Physical applications; for example,
many  $VSI$ and $CSI$ spacetimes are exact solutions in supergravity
and string theory (to all perturbative orders in the~string
tension) \cite{string}.

\subsubsection{Higher dimensional Kundt spacetimes:}

The generalized N-dimensional \emph{Kundt metric} (or simply Kundt metric \cite{kramer}) can be written \cite{Higher}
 \BE d s^2=2d u\left[d v+H(v,u,x^k)d u+W_{i}(v,u,x^k)d x^i\right]+
 g_{ij}(u,x^k)d x^id x^j. \label{Kundt}\EE The~metric functions
 $H$, $W_{i}$ and $g_{ij}$ satisfy the Einstein equations ($i,j = 2, ..., N-2$).
It is convenient to introduce the null frame

\BE {\bl}= d u, \\ {\bf n}= d v+Hd u+W_{ i}{{\bf m}}^{ i},
\\ {\bf m}^{ i}= m^{i}_{ j}d x^{j}, \EE
such that $g_{ij}=m^{k}_{\ i}m_{kj}$ and where $m^{i}_{\ j}$ can
be chosen to be in upper triangular form by an appropriate choice
of frame. The metric (\ref{Kundt}) possesses a null vector field
${\bl} = \frac{\partial}{\partial{v}}$ which is  geodesic,
non-expanding, shear-free and non-twisting; i.e., $L_{ij} \equiv
\ell_{i;j}=0$ \cite{Higher}. The Ricci rotation coefficients are
thus given by: \BE L_{ab} = \ell_{(a;b)} = L \ell_{a}\ell_{b} +
L_i(\ell_{a} m^{i}_{b} + \ell_{b} m^{i}_{a}).  \EE

\subsubsection{Covariantly constant null vector:}

In general, the generalized Kundt metric has the non-vanishing
Ricci rotation coefficients $L$ and $L_{i}$. From  $L =0$, we
obtain $H_{,v} =0$. By making use of the upper triangular form of
$m^{i}_{\ j}$, it follows that $L_i = 0$ implies $W_{i,v} = 0$
\cite{mcnutt}. The remaining transformations can be used to
further simplify the remaining non-trivial metric functions.

Thus, the aligned, repeated, null vector $\bl$ of (\ref{Kundt}) is a
null Killing vector (KV) if and only if $H_{v}=0$ and
$W_{i,v}=0$ (whence the
 metric no longer has any $v$
dependence).  Furthermore, it follows
that if $\bl$ is a null KV then it
is also covariantly constant.  Without any further restrictions, the higher
dimensional metrics
admitting a null KV have Ricci and Weyl type {\bf II}.
Therefore, the most general metric that admits a covariantly constant null vector (CCNV) is (\ref{Kundt})
with $H = H(u,x^k)$ and $W_{i} = W_{i}(u,x^k)$
\cite{TOrtin,CFHP}.

\subsection{Spacetimes with vanishing  scalar curvature invariants in higher dimensions}

The following Theorem was proven in \cite{Higher} (generalizing a
theorem in 4 dimensions \cite{VSI}):

\begin{theorem}
\label{main-theorem}

All curvature invariants of all orders vanish in an
$N$-dimensional Lorentzian spacetime if and only if there exists
an aligned non-expanding ($S_{ij}=0$), non-twisting ($A_{ij}=0$), shearfree
geodesic null direction $\ell^a$ along which the Riemann tensor
has negative boost order.

\end{theorem}

An analytical form of the conditions in this Theorem are as
follows: \BE R_{abcd} = 8 A_{i} \, \lnlm{i}+ 8 B_{ijk} \,
\mmlm{i}{j}{k}+ 8 C_{ij} \, \lmlm{i}{j} \label{RRR} \EE (i.e., the
Riemann tensor, and consequently the Weyl and Ricci tensors, are of
algebraic type {\bf III}, {\bf N} or {\bf O} \cite{class}), and
\BE \ell_{a ; b} = L_{11} \ell_a \ell_b  + L_{1i} \ell_a
\msub{i}{b} + L_{i1} \msub{i}{a} \ell_b; \label{l_ab-VSI} \EE
hence, $VSI$ spacetimes belong to the generalized~Kundt class
\cite{kramer}. It follows that any {$VSI$} metric can
be written in the generalized Kundt form  (\ref{Kundt}), where
local coordinates can be chosen so that the transverse metric is
flat; i.e., $g_{ij}=\delta_{ij}$ \cite{CSI}. The~metric functions $H$ and
$W_{i}$ in the Kundt metric (which can be obtained by substituting
$\sigma=0$ in eqns. (\ref{KundtCH}) and (\ref{KundtCW})), satisfy
the remaining vanishing scalar invariant conditions and
the~Einstein equations (which gives rise to two cases, which can
be characterized by $W^{(1)}_1 = -\frac{2}{x^1}\epsilon;\quad
W^{(1)}_i = 0,~ i\neq 1$, where $\epsilon =0$ or $1$).

Further progress can be made by classifing the $VSI$ metrics in
terms of their Weyl type ({\bf III}, {\bf N} or {\bf O}) and their
Ricci type {{\bf III}, {\bf  N} or {\bf O}), and the form of the
non-vanishing Ricci rotation coefficients $L_{ab}$. The Ricci
tensor can be written as \BE R_{ab} = \Phi \ell_{a}\ell_{b} +
\Phi_{i} (\ell_{a} m^{i}_{b} + \ell_{b} m^{i}_{a}). \label{ricci}
\EE The Ricci type is {\bf N} if $\Phi_{i}=0 = R_{1i}$ (otherwise
the Ricci type is {\bf III}; Ricci type {\bf O} is vacuum), whence
${H}^{(1)}(u,x^k)$ is determined in terms of the functions ${W}_{
i}^{(0)}(u,x^k)$. The remaining non-zero components can be simplified and
chosen to be constant by an appropriate choice of frame. The Weyl tensor can be expressed as \BE
C_{abcd} = 8 \Psi_{i} \ell_{\{a} n_b \ell_{c} m^{i}_{d\}} + 8
\Psi_{ijk} m^{i}_{\{a} m^{j}_b \ell_c m^{k}_{d\}} + 8 \Psi_{ij}
\ell_{\{a} m^{i}_b \ell_{c} m^{j}_{d\}} . \label{WeylIIIN} \EE The
case $\Psi_{ijk} \not= 0$ is of Weyl type {\bf III}, while
$\Psi_{ijk} = 0$ (and consequently also $\Psi_{i} = 0$)
corresponds to type {\bf  N}. Note that $\Psi_{ij}$ is symmetric
and traceless. $\Psi_{ijk}$ is antisymmetric in the first two
indices with $\Psi_i=2 \Psi_{ijj}$, and in vacuum also satisfies
$\Psi_{\{ijk\}}=0$. Further subclasses can be considered
\cite{class}. The subclass {\bf III}(a) is defined by
$C_{011(n+1)}=0$; this subcase includes the higher dimensional
{\em{generalized Kundt waves}} and the higher dimensional {\em
generalized pp-waves} \cite{RT,hortseyt}. Additional constraints
on $\Psi_{ij}$, $\Psi_{ijk}$, etc. can be obtained by employing
the Bianchi and Ricci identities \cite{Bianchi}. In addition, the
spatial tensors (indices $i,j$) can be written in canonical form
when the remaining coordinate and frame freedom is utilized. In
Table $1$ in \cite{CFHP}, all of the $VSI$ spacetimes supported by
appropriate bosonic fields were presented and the metric functions
explicitly listed.

As noted earlier, the aligned, repeated, null vector $\bl$ of (\ref{Kundt}) is a null 
KV and consequently covariantly constant if and only if $H_{,v}=0$ and
$W_{i,v}=0$ 
(these metric functions are defined in eqns. (\ref{KundtCH}) and (\ref{KundtCW})),
whence the metric no longer has any $v$ dependence.
Therefore, in higher dimensions a $VSI$ spacetime which admits a
CCNV is, in general, of Ricci type {\bf III} or {\bf N} and Weyl type {\bf
III} or {\bf N}. The subclass of Ricci type {\bf N} CCNV spacetimes are
related to the ($F=1$) chiral null models of \cite{hortseyt}. The
subclass of Ricci type {\bf N} and Weyl type {\bf III} (a)
spacetimes (in which the functions $W_{i}$ satisfy further
restrictions \cite{CFHP}) are related to the relativistic gyratons
\cite{frolov}. The subclass of Ricci type {\bf N} and Weyl type
{\bf N} spacetimes are the {\em generalized pp-wave spacetimes}.

\newpage

\subsection{Spacetimes with constant scalar curvature invariants in higher dimensions}

Lorentzian spacetimes for which all polynomial scalar invariants
constructed from the Riemann tensor and its covariant derivatives
are constant ($CSI$ spacetimes) were studied in \cite{CSI}. The
set of all locally homogeneous spacetimes is denoted by $H$.
Clearly $VSI \subset CSI$ and $H \subset CSI$. Let us denote by
${CSI_R}$ all reducible $CSI$ spacetimes that can be built from
$VSI$ and $H$ by (i) warped products (ii) fibered products, and
(iii) tensor sums. Let us denote by ${CSI_F}$ those spacetimes for
which there exists a frame with a null vector $\bl$ such that all
components of the Riemann tensor and its covariants derivatives in
this frame have the property that (i) all positive boost weight
components (with respect to $\bl$) are zero and (ii) all zero
boost weight components are constant. Finally, let us denote by
${CSI_K}$ those $CSI$ spacetimes that belong to the
higher-dimensional Kundt class; the so-called Kundt $CSI$
spacetimes. In particular, the relationship between ${CSI_R}$,
${CSI_F}$, ${CSI_K}$ and $CSI \backslash H$ was studied in
arbitrary dimensions (and considered in more detail in the
four-dimensional case) in \cite{CSI}. We note that by construction
${CSI_R}$ is at most of Weyl type {\bf II} (i.e., of type {\bf
II}, {\bf III}, {\bf N} or {\bf O} \cite{class}), and by
definition ${CSI_F}$ and ${CSI_K}$ are at most of Weyl type {\bf
II} (more precisely, at most of Riemann type {\bf II}). In
4 dimensions, ${CSI_R}$, ${CSI_F}$ and ${CSI_K}$ are closely
related, and it is plausible that $CSI \backslash H$ is at most of
Weyl type {\bf II}.

For a Riemannian manifold, every $CSI$ is locally
homogeneous $(CSI \equiv H)$ \cite{PTV}. This is not true for Lorentzian
manifolds.  However, for every $CSI$ spacetime with particular constant
invariants there is a homogeneous spacetime (not necessarily
unique) with precisely the same constant invariants.  This
suggests that $CSI$ can be constructed from $H$ and $VSI$ (e.g.,
${CSI_R}$). Indeed, from the work in \cite{CSI} it was conjectured
that if a spacetime is {$CSI$} then there exists a null frame in
which the Riemann tensor and its derivatives can be brought into
one of the following forms: (i) the Riemann tensor and its
derivatives are constant, in which case we have a locally
homogeneous space, or (ii) the Riemann tensor and its derivatives
are of boost order zero with constant boost weight zero components
at each order, which implies that the Riemann tensor (and hence the
Weyl tensor) is of type {\bf II}, {\bf III}, {\bf N} or {\bf O}
(the ${CSI_F}$ conjecture). This then suggests that if a spacetime
is {$CSI$}, the spacetime is either locally homogeneous or
belongs to the higher dimensional Kundt $CSI$ class (the ${CSI_K}$
conjecture) and that it can be constructed from locally
homogeneous spaces and $VSI$ spacetimes (the ${CSI_R}$
conjecture). The $CSI$ conjectures were proven in three dimensions
in \cite{CSI3d}. Partial results were obtained in four 
dimensions in \cite{CSI}.

In a higher dimensional Kundt spacetime, there exists a local
coordinate system in which the metric takes the form
(\ref{Kundt}). In \cite{CSI} it was shown that for Kundt $CSI$
(${CSI_K}$) spacetimes there exists (locally) a coordinate
transformation $(v',u',x'^i)=(v,u,f^i(u;x^k))$ that preserves the
form of the metric such that $\tilde{g}'_{ij,u'}=0$ (i.e.,
$\tilde{g}'_{ij}$ is independent of $u$) and that $d
s_{Hom}^2=\tilde{g}'_{ij}d x'^i d x'^j$ is a locally homogeneous
space. The remaining $CSI$ conditions then imply that \BE
W_{i}(v,u,x^k)=v{W}_{i}^{(1)}(u,x^k)+{W}_{i}^{(0)}(u,x^k),\label{KundtCH}
\EE \BE
H(v,u,x^k)=\frac{v^2}{8}\left[4\sigma+({W}_i^{(1)})({W}^{(1)i})\right]+
v{H}^{(1)}(u,x^k)+{H}^{(0)}(u,x^k), \label{KundtCW}\EE where
$\sigma$ is a constant \cite{CSI}. For $CSI$ spacetimes in which the metric
functions are independent of $v$ (i.e., 
${W}_i^{(1)} = {H}^{(1)} = \sigma=0$), the null vector $\bl$ is a CCNV; 
indeed, CCNV-$CSI$ spacetimes are necessarily Kundt and are
consequently of this form \cite{mcnutt}.
CCNV-$CSI$ spacetimes
were studied in \cite{CFH,mcnutt}.


\subsection{Supergravity and Supersymmetry}

The $VSI$ and $CSI$ spacetimes are of particular interest
since they are solutions of supergravity or superstring theory, when supported by
appropriate bosonic fields \cite{CFH,VSISUG}.
The supersymmetry properties of these
spacetimes have also been discussed.

\subsubsection{Discussion:}

In 4 dimensions it is known that pp-wave spacetimes, in which all
of the~scalar curvature invariants vanish \cite{jordan}, are exact
vacuum solutions to string theory to all orders in the string tension
scale $\alpha'$, even when the dilaton field and antisymmetric
tensor fields (which are also massless fields of string theory) are
included \cite{string}. It has been shown that in 4 dimensions all
of the $VSI$ spacetimes are classical solutions of the string
equations to all orders in $\sigma$-model perturbation theory by
showing that all higher order correction terms vanish
\cite{coley}.

The classical equations of motion for a metric in string theory
can be expressed in terms of $\sigma$-model perturbation theory
\cite{cfmp}, through the Ricci tensor $R_{\mu \nu}$ and higher
order corrections in powers of $\alpha'$ and terms constructed
from derivatives and higher powers of the Riemann curvature
tensor. The only non-zero symmetric second-rank tensor covariantly
constructed from scalar invariants and polynomials in
the~curvature and their covariant derivatives in $VSI$ spacetimes is
the Ricci tensor, and hence all higher-order terms in the~string
equations of motion automatically vanish \cite{string}. More
importantly,  other bosonic massless fields of the string can be
included. For example, a dilaton $\Phi$ and an
antisymmetric (massless field) $H_{\mu v \rho}$ can be
included. Assuming
appropriate forms for $\Phi$ and $H_{\mu v \rho}$ in $VSI$
spacetimes, where $H^2 = \nabla^2 \Phi = (\nabla \Phi)^2 =0$, when
the field equation
$$ R_{\mu \nu} - \frac{1}{4} H_{\mu \rho \sigma}H_\nu \; ^{\rho \sigma}
- 2 \nabla_\mu \nabla_\nu \Phi =0,  $$ is satisfied (typically
this equation constitutes a single differential equation for the
functions $\Phi$, $H_{\mu \nu \rho}$ and the metric functions),
all of the field equations  \cite{cfmp} are satisfied
to leading order in $\sigma$-model perturbation
theory (i.e., to order $\alpha^\prime$). Consequently, all higher
order corrections in $\sigma$-model perturbation theory, which are
of the form of second rank tensors and scalars constructed from
$\nabla_\mu \Phi$, $H_{\mu \nu \rho}$, the metric and their
derivatives, vanish for $VSI$ spacetimes. That is, $VSI$ spacetimes
are solutions to string theory to all orders in $\sigma$-model
perturbation theory \cite{coley}. It is plausible that a wide
class of $VSI$ solutions are exact solutions to string theory
non-perturbatively \cite{coley,string}.

In the context of string theory, it is of considerable interest to
study Lorentzian spacetimes in higher dimensions. 
String theory in higher dimensional generalizations of pp-wave  
backgrounds has been studied by many authors. 
It is known that such
spacetimes are exact solutions in string theory
\cite{string,hortseyt,RT}, and type-IIB superstrings in higher dimensional pp-wave
backgrounds were shown to be exactly solvable even in of
the~presence of the~RR five-form field strength \cite{RT}.
In particular,
supergravity theories have been studied in eleven and ten-dimensions, and
a class of 10-dimensional Ricci and Weyl
type {\bf N} pp-wave string spacetimes supported by non-constant
NS-NS  or R-R form fields (which depend on arbitrary
harmonic functions of the transverse  coordinates) were shown to
be exact type II superstring solutions to all orders in the string
tension \cite{RT}.
In addition,  a number of classical solutions of branes \cite{brane}
in higher dimensional pp-wave backgrounds have been studied in
order to better understand the non-perturbative dynamics of string
theories.

\subsubsection{VSI Supergravity theories:}

In \cite{VSISUG}
it was shown that the higher-dimensional 
$VSI$ spacetimes with fluxes and dilaton are solutions of type IIB
supergravity, and it was argued that they
are exact solutions in string theory.
$VSI$ solutions of IIB supergravity with NS-NS and
RR fluxes and dilaton have been constructed. The solutions are classified according to their
Ricci type ({\bf  N} or {\bf III}). The Ricci type {\bf  N} solutions are
generalizations of pp-wave type IIB supergravity solutions. The
Ricci type {\bf III} solutions are characterized by a non-constant
dilaton field.

In particular, it was shown that all Ricci type {\bf  N} $VSI$
spacetimes are solutions of supergravity (and it was argued that Ricci
type {\bf III} $VSI$ spacetimes are also supergravity solutions if
supported by appropriate sources). A number of new $VSI$ Ricci type
{\bf III} and Ricci type {\bf N} type IIB supergravity solutions
were presented explicitly. It was also argued that, in general,
the $VSI$ spacetimes are exact string solutions to all orders in the
string tension. As noted above, it is the higher dimensional
generalized pp-wave spacetimes, that are known to be exact
solutions in string theory \cite{string,hortseyt,RT}, that have
been most studied in the literature.

\subsubsection{CSI Supergravity theories:}

A number of $CSI$ spacetimes are also known to be
solutions of supergravity theory when supported by
appropriate bosonic fields \cite{CFH}.
It is known that $AdS_d \times S^{(N-d)}$ (in short $AdS\times S$)
is an exact solution of supergravity (and preserves the maximal
number of supersymmetries) for certain values of (d,N) and for
particular ratios of the radii of curvature of the two space
forms. Such spacetimes  (with $d=5,N=10$) are supersymmetric
solutions of IIB supergravity (and there are analogous solutions
in $N=11$ supergravity) \cite{kumar}.  $AdS \times S$ is an
example of a $CSI$ spacetime \cite{CSI}. There are a number of other
$CSI$ spacetimes known to be solutions of supergravity and admit
supersymmetries; namely, generalizations of $AdS \times S$ (for
example, see \cite{Gauntlett}) and (generalizations of) the chiral
null models \cite{hortseyt}. The $AdS$ gyraton (which is a $CSI$
spacetime with the same curvature invariants as pure $AdS$)
\cite{frolov} is a solution of gauged supergravity  \cite{Caldarelli}
(the $AdS$ gyraton can be cast in the Kundt form \cite{CFH}).

More general $CSI$ spacetimes can be investigated to determine whether they are
solutions of supergravity. For example, product manifolds of
the form $M\times K$ can be considered, where $M$ is an Einstein
space with negative constant curvature and $K$ is a (compact)
Einstein-Sasaki spacetime (in \cite{membranes} supersymmetric
solutions of $N=11$ supergravity, where $M$ is the squashed $S^7$,
were given). The warped product of $AdS_3$ with an 8-dimensional
compact (Einstein-Kahler) space $M_8$ with non-vanishing 4-form
flux are supersymmetric solutions of N=11 supergravity
\cite{Gauntlett}.

A class of supergravity $CSI$ solutions have been built from a $VSI$
seed and locally homogeneous (Einstein) spaces by warped products,
fibered products and tensor sums \cite{CSI}, yielding
generalizations of $AdS\times S$ or $AdS$ gyratons \cite{CFH}. In
particular, solutions obtained by restricting attention to
Ricci type {\bf N} CCNV spacetimes were considered. Some explicit
examples of $CSI$ supergravity spacetimes were constructed by taking
a homogeneous (Einstein) spacetime,
$(\mathcal{M}_{Hom},\tilde{g})$, of Kundt form and generalising to an
inhomogeneous spacetime, $(\mathcal{M},{g})$,  by including
arbitrary Kundt metric functions (by construction, the curvature
invariants of $(\mathcal{M},{g})$ will be identical to those of
$(\mathcal{M}_{Hom},\tilde{g})$); a number of 5-dimensional
examples were given explicitly, in which $ds_{Hom}^2$ was taken to
be Euclidean space or hyperbolic space \cite{cardoso}.

\subsubsection{Supersymmetry:}

The supersymmetric properties of $VSI$ and $CSI$ spacetimes have also
been studied. It is known that in general if a spacetime admits a
Killing spinor, it necessarily admits a null or timelike Killing
vector (KV). Therefore, a necessary (but not sufficient) condition for
a particular supergravity solution to preserve some supersymmetry
is that the spacetime possess such a KV. The existence of null or timelike
KVs in $VSI$ spacetimes was studied in \cite{VSISUG}.
Subsequently, the
supersymmetry properties of $VSI$ type IIB supergravity solutions
admitting a CCNV were investigated; in particular, 
the previously studied Weyl type {\bf  N} spacetimes were
discussed extensively and new explicit examples of Weyl type {\bf
III}(a) NS-NS (one-half) supersymmetric solutions  (when the axion
and metric functions are appropriately related) were presented 
\cite{VSISUG}. Supersymmetry has also been studied in $CSI$
supergravity solutions, particularly in the CCNV subclass of
$CSI$ spacetimes \cite{CFH}. 

\newpage

\section{Conclusions and Outlook}

We have reviewed the algebraic classification of the Weyl tensor in
higher dimensional Lorentzian manifolds. This classification is achieved  by
characterizing algebraically special Weyl tensors by means of the
existence of aligned null vectors of various orders of alignment.
In most applications 
this classification suffices and the details of a more complete
classification are not necessary. However,
further classification can be obtained using reducibility and by determining the alignment
type.

One outstanding issue is that it would be very useful
to be able to find a more practical way of determining the
Weyl type of a spacetime such as, for example, by employing
higher dimensional scalar invariants. 
In particular, it would be useful to find
necessary conditions for a type  {\bf I} or
type  {\bf II} spacetime in terms of simple invariants.

We have discussed a number of applications of this classification scheme,
used in conjunction with the higher
dimensional frame formalism developed. In future work it would be important to prove
higher-dimensional versions of the Goldberg-Sachs theorem and the
peeling theorem. It would also be of interest to study the five dimensional case
and type {\bf D} spacetimes in more detail.

We also discussed the higher dimensional 
Lorentzian spacetimes with vanishing  scalar curvature
invariants ($VSI$ spacetimes)  and constant scalar
curvature invariants ($CSI$ spacetimes). 
It has been conjectured that if a spacetime
is {$CSI$}, then the spacetime is either locally homogeneous or
belongs to the higher dimensional Kundt class and can be constructed from locally
homogeneous spaces and $VSI$ spacetimes. It remains to prove the various $CSI$
conjectures, particularly in the case of four 
dimensions. It is also of importance to further study 
supergravity and supersymmetry in
$VSI$ and $CSI$ spacetimes.

{\em Acknowledgements}. This review is based on work done in collaboration with R.
Milson, V. Pravda, A. Pravdov\' a, N. Pelavas and S. Hervik, and
I would like to thank my collaborators for helpful comments on the current manuscript.
This work was supported by
NSERC of Canada.

\end{document}